\documentclass[runningheads,a4paper]{llncs}

\usepackage{graphicx}
\usepackage{algorithmic}
\usepackage[ruled,vlined]{algorithm2e}
\usepackage{url}
\usepackage{alltt}
\urldef{\mailsa}\path|pat@dcs.gla.ac.uk|

\newlength{\halftextwidth}
\usepackage{a4wide}
\usepackage{multicol}
\usepackage{listings}
\usepackage{amsopn}
\DeclareMathOperator{\mod}{mod}

\begin{document}

\mainmatter  

\title{Distributing an Exact Algorithm for Maximum Clique: maximising the costup\\
TR-2012-334}

\titlerunning{Distributing an Exact Algorithm for Maximum Clique}

\author{Ciaran McCreesh and Patrick Prosser}

\authorrunning{McCreesh and Prosser}

\institute{School of Computing Science,\\
  University of Glasgow, Glasgow, Scotland\\
pat@dcs.gla.ac.uk\\}

\maketitle

\lstset{ %
language=pascal,              
basicstyle=\scriptsize,     
numbers=left,               
numberstyle=\scriptsize,    
numbersep=10pt,             
frame=trBL,                 
captionpos=b,               
breaklines=true,            
breakatwhitespace=false,    
showstringspaces=false,     
}
\begin{abstract}
We take an existing implementation of an algorithm for the maximum clique
problem and modify it so that we can distribute it over an ad-hoc cluster of machines. 
Our goal was to achieve a significant 
speedup in performance with minimal development effort, i.e. a \emph{maximum costup}. We present a simple 
modification to a state-of-the-art exact algorithm for maximum clique that allows us to distribute it across many machines. 
An empirical study over large hard benchmarks shows that speedups 
of an order of magnitude are routine for 25 or more machines. 
\end{abstract}

\section{Introduction}
\label{sec:intro}
\vspace{-1.5mm}
Intel's tera-scale computing vision (\emph{a parallel path to the future})
is to aim for hundreds of cores on a chip. 
In their white paper \cite{intel2006} Intel puts ``programmability'' at the top of the list of challenges 
for the multi-core era, and considers the development of multi-core software to be amongst the greatest 
challenges for tera-scale computing. However, Hill and Marty \cite{hillMarty}
propose that in exploiting multi-cores we should not just aim for \emph{speedup} but also \emph{costup}, i.e. an increase
in performance that is greater than the increase in cost, be it measured in money or energy.

And that is our target, to maximise costup. We were presented with the following challenge. The second author had just completed an 
empirical study of exact algorithms for the maximum clique problem and a variety of algorithms had been implemented
in Java \cite{tr-2012-333}. Being summer, a University environment and the students away, we had access to over 100 teaching machines that we could
use but not modify (limiting us to using SSH, NFS and Java). We didn't have a shared memory system with hundreds of cores but we
did have access to a hundred networked PCs, but only for four weeks.
Could we take one of our programs, make minimal changes to it, distribute it over the machines 
available and solve some big hard problems quickly? To be more precise, starting mid-week (Wednesday 4th July) 
could we do this by Friday (the 6th),
use the available resources over the weekend (7th and 8th), solve some big hard problems and analyse the results on Monday (the 9th)? 
The problem studied was the \emph{maximum clique problem}.

\begin{definition}
A \textbf{simple undirected graph} G is a pair (V(G),E(G)) where V(G) is a set of vertices and E(G) a set of edges. An edge
$\{u,v\}$ is in E(G) if and only if $\{u,v\} \subseteq$ V(G) and vertex $u$ is adjacent to vertex $v$.
\end{definition}

\begin{definition}
 A \textbf{clique} is a
set of vertices C $\subseteq$ V(G) such that every pair of vertices in C is adjacent in G.
\end{definition}

Clique is one of the six basic NP-complete problems given in \cite{gareyJohnson}. It is posed as a decision
problem [GT19]: given a simple undirected graph G = (V(G),E(G)) and a positive integer k $\leq$ $|$V(G)$|$ does G contain a clique of size
k or more? The optimization problems is then to find a \emph{maximum clique}, whose size is denoted $\omega$(G).
A graph can be \emph{coloured} by assigning colour values to vertices such that adjacent vertices take different colour values.
The minimum number of different colours required is then the
\emph{chromatic number} of the graph $\chi$(G), and $\omega$(G) $\leq \chi$(G). 
Therefore a colouring of the graph G can be used as an upper bound on $\omega$(G).
Finding the chromatic number is NP-hard, but fast approximations exist such as \cite{welshPowell,brelaz}.

In the next section we describe our starting point, the algorithm MC. We then show a simple modification to the algorithm
that allows distribution at various levels of granularity. Implementation details are then given followed by a 
report of our computational study. We then reflect, considering what we might have done differently given more time, and then conclude.

\section{An Exact Algorithm for Maximum Clique (MC)}
\label{sec:mcBin}
\vspace{-1.5mm}
We can address the optimization problem with an exact algorithm,
such as a backtracking search
\cite{fahle,regin2003,wood97,carraghanPardalos90,pardalosRodgers92,prjo2002,segundo2011,segundo2011b,Konc_Janezic_2007,tomita2003,tomita2010,aaai2010,carmoZuge}.
Backtracking search incrementally constructs the set C (initially empty) by choosing a \emph{candidate vertex}
from  the \emph{candidate set} P (initially all of the vertices in V(G)) and then adding it to C.
Having chosen a vertex the candidate set is then updated, removing vertices
that cannot participate in the evolving clique. If the candidate set is empty then C is maximal (and if it is a maximum we save it)
and we then backtrack. Otherwise P is not empty and we continue our search, selecting from P and adding to C.

There are other scenarios where we can cut off search, e.g. if what is in P is insufficient
to unseat the champion (the largest clique found so far) then search can be abandoned. That is, an upper bound can be computed.
Graph colouring can be used to compute an upper
bound during search, i.e. if the candidate set can be coloured with $k$ colours then it can contain a clique no larger than $k$
\cite{wood97,fahle,segundo2011,Konc_Janezic_2007,tomita2003,tomita2010}. There are also heuristics that can be used when selecting
the candidate vertex, different styles of search, different algorithms to colour the graph and different
orders in which to do this.

For our study we use MC ({\bf M}aximum {\bf C}lique), Algorithm \ref{MC}. MC is essentially algorithm MCSa1 in \cite{tr-2012-333}
and corresponds to Tomita's MCS \cite{tomita2010} with the colour 
repair step removed. MCSa1 is a state of the art
algorithm and a close competitor to San Segundo's BBMC \cite{segundo2011}\footnote{A java implementation
of MCQ, MCR, MCSa, MCSb and BBMC is available at http://www.dcs.gla.ac.uk/$\sim$pat/maxClique}.
The algorithm performs a binomial search
(see pages 6 and 7 of \cite{dek}) and uses a colour cutoff. Vertices are selected from the candidate set P 
and added to the growing clique C. The graph induced by the vertices in P is coloured such that each vertex in P has an associated colour.
Vertices are then selected in non-increasing colour order (largest colour first) for inclusion in C. 
Assume a vertex v is selected from P and has colour k.
The graph induced by the vertices in P, including v, can be coloured with k colours and 
can therefore contain a clique no larger than k. Consequently if the cardinality of C plus k is no larger than
that of the largest clique found so far search can be abandoned. Crucial to the success of the algorithm is the quality of
the colouring of the candidate set and the time taken to perform that colouring. Empirical evidence
suggests that good performance can be had with any of the three colour orderings studied in \cite{tr-2012-333}.

Algorithms 1 and 2 are presented as procedures that deliver a result, possibly void (line 12) and
possibly a tuple (line 22). We assume that a {\bf Set} is an order preserving structure such that when an item $v$
is added to a {\bf Set} $S$, i.e. $S \gets S \cup \{v\}$, the last element in $S$ will be $v$ and when
$S_{2} \gets S_{0} \cap S_{1}$ the elements in $S_{2}$ will occur in the same order as they appear in $S_{0}$.

MC, Algorithm \ref{MC}, takes as parameter a graph $G$ and has three global variables (lines 3 to 5): integer $n$ the 
number of vertices in $G$, $C_{max}$ the set of vertices in 
the largest maximal clique found so far and integer $\omega_{*}$ the size of that clique. MC then calls expand (line 8) to explore the 
backtrack tree to find a largest clique in $G$. Procedure $expand$ is called (in line 8) with three arguments: the candidate set $P$ (line 6) 
and the growing clique $C$ (line 7, initially empty) and the graph $G$. Initially the candidate set contains all the vertices in the graph, V(G), 
and is sorted in non-increasing degree order (the call to $sort$ in line 6) and this order is then used for colouring the graph induced by $P$.

Procedure $expand$ starts by colouring the graph induced by $P$ (step 12), delivering a pair $(S,colour)$ where $S$ is a stack
of vertices and colour is an array of colours (integers). If vertex $v$ is at the top of the stack then vertex $v$ has colour $colour[v]$ and 
all vertices in the stack have a colour less than or equal to $colour[v]$. The procedure iterates over the stack (while loop of line 13),
selecting and removing a vertex from the top of the stack (line 14). If the colour of that vertex is too small then
the graph induced by $v$ and the remaining vertices in the stack and the vertices in the growing clique $C$
will be insufficient to unseat the current champion and search can be terminated (line 15). Otherwise the vertex $v$ is added to
the clique (line 16) and a new candidate set is produced $P'$ (line 17) where $P'$ is the set of vertices in $P$ that are
adjacent to the current vertex $v$ (where $N(v,G)$ delivers the \emph{neighbourhood} of $v$ in $G$), 
consequently each vertex in $P'$ is adjacent to all vertices in $C$. If the new candidate set is empty then $C$ is maximal
and if it is larger than the largest clique found so far it is saved (line 18). But if $P'$ is not empty $C$ is not maximal
and $C$ can grow via the recursive call to $expand$ (line 19). Regardless, when all possibilities of expanding the 
current clique with the vertex $v$ have been considered that vertex can be removed from the current clique (line 20) 
and from the candidate set (line 21).

Procedure $colourSort$, line 22, corresponds to Tomita's NUMBER-SORT in \cite{tomita2003}. The vertices in $P$ are sequentially 
coloured (assigned a 
colour number) and sorted (the vertices are delivered as a stack such that the vertices in the stack appear in
descending colour order, largest colour at top of stack). In line 24 an integer array of colours is created and in
line 25 an array of sets is produced, such that the set $colourClass[k]$ contains non-adjacent vertices that have colour $k$, i.e.
$colourClass[k]$ is an \emph{independent set}. The candidate set is iterated over in line 28, and this will be in non-increasing
degree order as a consequence of the initial sorting in line 6. Line 30 searches for a colour class for the current vertex $v$:
if any vertex in $colourClass[k]$ is adjacent to $v$ we then look in the next colour class (increment $k$)\footnote{$adjacent$ 
(lines 37 to 41) delivers $true$ if in the graph $G$  vertex $v$ is adjacent to a vertex in the set $S$.}.
In line 31 we have found a suitable $colourClass[k]$ (possibly empty) and we add $v$ to that $colorClass$, assign that colour to the vertex
(line 32) and take note of the number of colours 
used (line 33)\footnote{Lines 28 to 33 correspond to the sequential colouring of \cite{welshPowell}.}.  
Once all vertices have been added to colour classes 
we sort the vertices using a pigeonhole sort (lines 34 and 35): for each colour class we push the vertices in that
colour class onto the stack. The procedure then finishes by returning the colour-sorted vertices with their colours (line 36) as a pair.

\begin{algorithm}
\DontPrintSemicolon
\nl $\textbf{Set} ~ MC(\textbf{Graph} ~ G)$ \;
\nl \Begin{
\nl $\textbf{Global} ~ n ~ \gets |V(G)|$ \;
\nl $\textbf{Global} ~ C_{max} ~ \gets \emptyset$ \;
\nl $\textbf{Global} ~ \omega_{*} \gets 0$ \;
\nl $P \gets sort(V(G),G)$ \;
\nl $C \gets \emptyset$ \;
\nl $expand(C,P,G)$ \;
\nl $\textbf{return} ~ C_{max}$ \;
}
\;
\nl $\textbf{void}~expand(\textbf{Set}~C,\textbf{Set}~P,\textbf{Graph}~G)$ \;
\nl \Begin{
\nl $(S,colour) \gets colourSort(P,G)$ \;
\nl \While {$S \neq \emptyset$}{
\nl $v \gets pop(S)$ \;
\nl \lIf {$colour[v] + |C| \leq |C_{max}|$}{$\textbf{return}$} \;
\nl $C \gets C \cup \{v\}$ \;
\nl $P' \gets P \cap N(v,G)$ \;
\nl \lIf {$P' = \emptyset~\bf{and}~|C| > \omega_{*}$}{$C_{max} \gets C, \omega_{*} \gets |C|$} \;
\nl \lIf {$P' \neq \emptyset$}{$expand(C,P',G)$} \;
\nl $C \gets C \setminus \{v\}$ \;
\nl $P \gets P \setminus \{v\}$ \;
 }
}
\;
\nl $(\textbf{Stack},\textbf{integer} [])~colourSort(\textbf{Set}~P,\textbf{Graph}~G)$ \;
\nl \Begin{
\nl $colour \gets~new~\textbf{integer} [n]$ \;
\nl $colourClass \gets~new~\textbf{Set}[n]$ \;
\nl $coloursUsed \gets 0$ \;
\nl $S \gets ~new~\textbf{Stack}(\emptyset)$ \;
\nl \For {$v \in P$}{
\nl $k \gets 1$\;
\nl \lWhile {$adjacent(v,colourClass[k],G)$}{$k \gets k + 1$}\;
\nl $colourClass[k] \gets colourClass[k] \cup \{v\}$ \;
\nl $colour[v] \gets k$ \;
\nl $coloursUsed \gets max(k,coloursUsed)$\;
 }
\nl \For {$i \gets ~ 1 ~ to~ coloursUsed$}{
\nl \lFor {$v \in colourClass[i]$}{$push(S,v$)}
 }
\nl $\textbf{return}~(S,colour)$ \;
}
\;
\nl $\textbf{boolean}~adjacent(\textbf{integer}~v,\textbf{Set}~S,\textbf{Graph}~G)$ \;
\nl \Begin{
\nl \For {$w \in S$} { 
\nl \lIf {$adjacent(v,w,G)$}{$\textbf{return}~true$} \;
 }
\nl $\textbf{return}~false$ \;
}
\caption{The sequential maximum clique algorithm MC}
\label{MC}
\end{algorithm}

\section{Distributing MC (MCDist)}
\label{sec:models}
\vspace{-1.5mm}
There are a number of ways we might distribute MC across $p$ processors. We could split the problem into $p$ parts, 
run each part individually and merge the results, as a map-reduce style implementation. But we cannot partition the 
problem into $p$ roughly-equally sized chunks. We could instead split the problem into $n$ parts, where $n$ is the 
number of vertices in the graph, and then use a worker pool model of execution. Each job would expand a backtrack 
tree rooted on a node at level 1 where the current clique contained
a single vertex. That is, we kick off $n$ jobs each with a different clique of size one. 
But given the uneven size of the search trees, $n$ is likely still too small to give well-balanced workloads. More generally, 
we could divide at level 2 potentially
kicking off $n \choose 2$ jobs where each process has an initial clique containing two adjacent vertices.
More generally we might kick off $m$ jobs where each job expands ${n \choose k}/m$ backtrack trees
rooted at specified nodes at depth $k$. MCDist allows us to do that.

In Algorithm \ref{MCDist} MCDist (line 1) takes the 
following arguments: the graph $G$, a set of sets $T$ where 
each element of $T$ is of size $arity$, and integer $c$ the size of the largest clique reported by other processes. Elements
of the set $T$ describe the nodes in the backtrack tree to be expanded by this process. For example, if 
$T = \{\{1\}\}$ arity would equal 1 and a call to $MCDist(G,T,1,0)$ would explore the backtrack tree rooted on the clique $\{1\}$.
If $T = \{\{1\},\{2\},\ldots,\{n\}\}$ a call to $MCDist(G,T,1,0)$ would be equivalent to the call 
$MC(G)$\footnote{... as would a call to $MCDist(G,\{\emptyset\},n+1,0)$.}.
If $T = \{\{1,2,3\},\{1,2,4\},\ldots,\{1,2,n\},\{1,3,4\},\ldots,\{n-2,n-1,n\}\}$ with $arity = 3$ and $c = k$
$MCDist$ would start expanding search from all triangles (level 3) looking for cliques of size greater than $k$.
And finally, we might divide the set of edges $E(G)$ equally amongst the sets $T_1$ to $T_m$ and
distribute calls to $MCDist(G,T_{i},2,c)$, for $i \in \{1,\ldots,m\}$, across the available processors. Each call to
$MCDist(G,T_{i},2,c)$ would occur within a separate job and the current best clique size $c$ would be used at dispatch time.

In Algorithm \ref{MCDist} we replace expand with its distributed counterpart distExpand. The essential difference between 
$expand$ and $distExpand$ is the call to $considerBranch$ in line 16, i.e. the lines 16 to 20 
in Algorithm \ref{MC} are now executed conditionally. Procedure $considerBranch$ determines if the current clique can be 
considered for expansion. If the clique is below or above the critical size then expansion can proceed, otherwise
search can proceed only if $|C| = arity$ and $C$ corresponds to a specified node of interest, i.e. $C \in T$.

\begin{algorithm}
\DontPrintSemicolon
\nl $\textbf{Set} ~ MCDist(\textbf{Graph}~G,\textbf{Set}~T,\textbf{integer}~arity,\textbf{integer}~c)$ \;
\nl \Begin{
\nl $\textbf{Global} ~ n ~ \gets |V(G)|$ \;
\nl $\textbf{Global} ~ C_{max} \gets \emptyset$ \;
\nl $\textbf{Global} ~ \omega_{*} \gets c$ \;
\nl $P \gets sort(V(G),G)$ \;
\nl $C \gets \emptyset$ \;
\nl $distExpand(C,P,T,arity,G)$ \;
\nl $\textbf{return} ~ C_{max}$ \;
 }
\;
\nl $\textbf{void} ~ distExpand(\textbf{Set}~C,\textbf{Set}~P,\textbf{Set}~T,\textbf{integer}~arity,\textbf{Graph}~G)$ \;
\nl \Begin{
\nl $(S,colour) \gets colourSort(P,G)$ \;
\nl \While {$S \neq \emptyset$}{
\nl $v \gets pop(S)$ \;
\nl \lIf {$colour[v] + |C| \leq |C_{max}|$} {$\textbf{return}$} \;
\nl \If {$considerBranch(C,T,arity)$}{
\nl $C \gets C \cup \{v\}$ \;
\nl $P' \gets P  \cap N(v,G)$ \;
\nl \lIf {$P' = \emptyset ~ \bf{and} |C| > \omega_{*}$}{$C_{max} \gets C,~ \omega_{*} \gets |C|$} \;
\nl \lIf {$P' \neq \emptyset$}{$distExpand(C,P',T,arity,G)$} \;
\nl $C \gets C \setminus \{v\}$ \;
  }
\nl $P \gets P \setminus \{v\}$ \;
 }
}
\;
\nl $\textbf{boolean} ~ considerBranch(\textbf{Set}~C,\textbf{Set}~T,\textbf{integer}~arity)$ \;
\nl \Begin{
\nl $\textbf{return} ~ |C| < arity ~ \textbf{or} ~ |C| > arity ~ \textbf{or} ~ C \in T$ \;
}
\caption{The distributed maximum clique algorithm MCDist}
\label{MCDist}
\end{algorithm}

Procedure $distExpand$ is similar to the search state re-computation technique used by \cite{perron1999} and it 
can be made more efficient. Assume the argument $arity$ equals $\alpha$. A call to $colourSort$ is made on each call to $distExpand$
as the clique $\{v_{1},\ldots,v_{\alpha-1}\}$ is incrementally constructed (repeatedly passing the test $|C| < arity$ at line 25) 
although it might ultimately be rejected when $C = \{v_{1},\ldots,v_{\alpha}\}$
(failing the test $C \in T$ at line 25). In the worst case, if $T$ was empty or contained a single tuple that did not correspond 
to any node in the backtrack tree, $colourSort$ would be called $O(\sum_{k=1}^{\alpha}{n \choose k})$ times to no effect. 
This is the cost of making a simple modification to $MC$ to give us $MCDist$. However in practice much of this cost is easily avoidable and
in our studies this overhead has always been tolerable. A second improvement is to enhance $considerBranch$ such that rather than 
delivering true if $|C| <  arity$ we deliver true if $|C| < arity$ and there exist a set $S \in T$ such that $C \subset S$. 
This will reduce redundant search. A further improvement is to remove elements of $T$ after they are expanded and put a test immediately
after line 11 that makes a {\bf return} if $T$ is empty.

\section{Implementation}
Possibilities for implementation were constrained by the available resources.
We intended to reuse an existing implementation of the algorithm, which was
written in Java. This was convenient, since the available machines all had a
JVM installed. For communication we were limited to NFS (with file-level
locking only) and SSH.

The existing code was modified in line with the differences between MC and
MCDist. Rather than passing $T$ explicitly, for a graph with $n$ vertices an integer
$t$ between $0$ and $8n-1$ was used to parameterise subproblems by splitting on
the second level of the binomial search tree, as if $arity$ was $2$.

A simple implementation is evident: we may number the top level nodes from
$n-1$ down to $0$, from right to left (corresponding to the order in which they
are popped from $S$). Then for a given $t$, every element of $T$ contains the
element we numbered $t \mod n$. On the second level of the tree, we again label
elements of $S$ from right to left, from $|S|-1$ down to $0$, and for a given
$t$, take those where the node's number divided by $n$ equals $t$, modulo $8$.

For each value of $t$, a file named for that number was created in a
``pending'' directory. These files were split by the last two digits between
100 subdirectories, to reduce contention and to keep each directory
sufficiently small to avoid NFS scalability problems (initially no split was
used, and lock contention prevented scalability beyond 10 machines).

The restriction to $8n-1$ jobs was necessary to avoid having too many files, but still have 
significantly more jobs than machines so that they can be distributed using a worker-pool model.
In addition a global ``best so far'' file was used to hold the value $c$ for MCDist. This
was set to contain $0$ initially.

The worker programs were launched by SSH (using public key / keychain logins to
avoid having to repeatedly enter passwords), with one worker program per
machine.  Each worker program reads in the graph, performs the initial colour
ordering on the vertices, and then starts running subproblems. The worker
randomly shuffles the 100 directories (to reduce contention), and then for each
directory in turn, while that directory is not empty, picks a job file from
that directory and moves it to a ``running'' directory. The worker then reads
in the ``best so far'', runs the subproblem, saves the result to the job file
and moves it to a ``results'' directory. The ``best so far'' is then updated,
and another problem is attempted.
Note that the ``best so far'' is only read in before starting any individual
subproblem.

Two sets of locking are required to avoid race conditions. Firstly, a lock file
is associated with each subproblem directory, to ensure that two machines
cannot both start running the same problem. Here exclusive locking is used.
Secondly, the ``best so far'' file needs to be locked to avoid races when it is
being updating, and to avoid the possibility of inconsistent reads. Initially,
a shared lock was used when reading in the value before launching a problem,
and an exclusive lock was used when checking and updating the file afterwards.
It was observed that this was a serious limiting factor on scalability (on
smaller problems, more than 50\% of the total runtime was being wasted waiting
for a lock). Thus, a more sophisticated mechanism for updating the file was
introduced.

The value in the ``best so far'' file may only increase over
time and its largest possible value is typically much smaller than the
number of subproblems. This means most of the exclusive locks we were obtaining
were in fact not being used to change the value. This suggests a better
strategy: when a worker finishes a problem, it obtains a shared lock on the
``best so far'' file, and compares the existing value to the value it
calculated. The lock is then released. If the newly calculated value is better
than the existing value, then an exclusive lock is obtained (avoiding the
possibility of deadlock, since the shared lock is already released). The value
is then re-compared (in case it has been updated in between releasing the shared
lock and obtaining the exclusive lock) and written if necessary.
We observed that in practice, this mechanism substantially reduced overhead.

After execution, obtaining the results is a simple matter of checking every
file in the ``results'' directory. (The ``best so far'' file only contains the
size of the best clique, not its members.) The existing implementation of the
algorithm produced data on the number of nodes (i.e. calls to $distExpand$) and the time spent working as
well as the clique found as part of its results.

\section{Computational Study}
\label{sec:study}
\vspace{-1.5mm}
Our study was performed over hard DIMACS\footnote{Available from ftp://dimacs.rutgers.edu/pub/dsj/clique} instances,
instances from the BHOSLIB suite 
(Benchmarks with Hidden Optimum Solutions\footnote{Available from http://www.nisde.buaa.edu.cn/$\sim$kexu/benchmarks/graph-benchmarks.htm})
and Erd\'{o}s-R\"{e}nyi random graphs.
These instances were selected because they took between minutes and weeks on a single machine and were hard enough to cover the start up
costs of distribution. Approximately 100 student lab PCs were used, running Fedora 13 with an AMD Athlon 64 X2
5200+, 4GBytes RAM and access to an NFS server. Machines were 
largely idle, but there was no guarantee regarding availability (several
machines were switched on and off or became unavailable whilst experiments were being run).
In all cases a problem was split into $8n$ jobs and these were then distributed across the machines, e.g.
frb35-17-1 has 450 vertices, 3,600 jobs were produced and these were dispatched over 25 machines, then 50 machines and finally 100 machines.
Run time was measured in seconds and is the difference between 
the wall clock time at the start of the first job and the wall clock time at the end of the last job. Clocks on the machines
were loosely synchronised, sometimes differing by a couple of seconds. The single machine runtimes are for the sequential (undistributed) 
algorithm and exclude the read-in times for the problem instance as in \cite{tr-2012-333}, whereas the distributed results include 
program startup and read-in times.

\subsection{DIMACS and BHOSLIB}
Table \ref{bigTable} shows the run time in seconds to find and prove optimality
using 25, 50 and 100 machines compared to the undistributed time. 
We list the number of vertices in the graph ($n$), the size of the maximum clique ($\omega$) 
and when more than one machine was
used the speed up (gain). The instances MANN-a45, brock400 and
p-hat500-3 are from DIMACS and frb* from BHOSLIB.

\begin{table}
\begin{center}
\begin{scriptsize}
\begin{tabular}{|c|c|c|c|c c|c c|c c|} \hline 
instance & $n$ & $\omega$ & $mc_{1}$  & $mc_{25}$ & (gain)  & $mc_{50}$ & (gain)   & $mc_{100}$  & (gain)   \\ \hline
MANN-a45     & 1,035 & 345   & 10,757      & 992      & (10.8)   & 1,009     & (10.7)   & 989 & (10.9) \\
brock400-1   & 400 & 27    & 4,973       & 186      & (26.7)   & 254      & (19.6)   &  121 & (41.1) \\
brock400-2   & 400 & 29    & 3,177       & 243      & (13.1)   & 137      & (23.2)   &  130 & (24.4) \\
brock400-3   & 400 & 31    & 2,392       & 113      & (21.2)   & 128      & (18.7)   &  121 & (19.8) \\
brock400-4   & 400 & 33    & 1,201       & 150      & (8.0)    & 122      & (9.8)    &  91 & (13.2) \\
frb30-15-1   & 450 & 30    & 11,430      & 679      & (16.8)   & 420      & (27.2)   & 247         & (46.3)   \\
frb30-15-2   & 450 & 30    & 17,649      & 1,046    & (16.9)   & 469      & (37.6)   & 427        & (41.3)   \\
frb30-15-3   & 450 & 30    & 5,877       & 442      & (13.3)   & 389      & (15.1)   & 192        & (30.6)   \\
frb30-15-4   & 450 & 30    & 31,176      & 787      & (39.6)   & 707      & (44.1)   & 701        & (44.5)   \\
frb30-15-5   & 450 & 30    & 9,827       & 620      & (15.9)   & 421      & (23.3)   & 1,131       & (8.7)    \\
frb35-17-1   & 595 & 35  & 624,722     & 32,106    & (19.5)   & 16,800    & (37.2)   & 33,978    & (18.4)   \\
frb35-17-2   & 595 & 35  & 1,133,097    & 69,105    & (16.4)   & 28,285    & (40.1)   & 27,978    & (40.5)   \\
frb35-17-3   & 595 & 35  & 331,542     & 26,898    & (12.3)   & 16,035    & (20.7)   & 12,677    & (26.2)   \\
frb35-17-4   & 595 & 35  & 359,966     & 29,255    & (12.3)   & 18,234    & (19.7)   & 16,759    & (21.5)   \\
frb35-17-5   & 595 & 35  & 1,921,917    & 100,715   & (19.1)   & 75,871    & (25.3)   & 40,144    & (47.9)   \\
p-hat500-3   & 500 & 50  & 3,051       & 112      & (27.2)   & 875      & (3.5)    & 119      & (25.6)   \\ \hline

\end{tabular}
\end{scriptsize}
\end{center}
\caption{Large hard instance: run time in seconds, using 1 to 100 machines. An entry of --- corresponds to job that did not terminate after one week.}
\label{bigTable}
\end{table}

\begin{figure}
\vspace{-4.5cm}
\begin{center}
\hspace{-1.5cm}
\begin{minipage}[t]{0.30\textwidth}
\includegraphics[height=9.0cm]{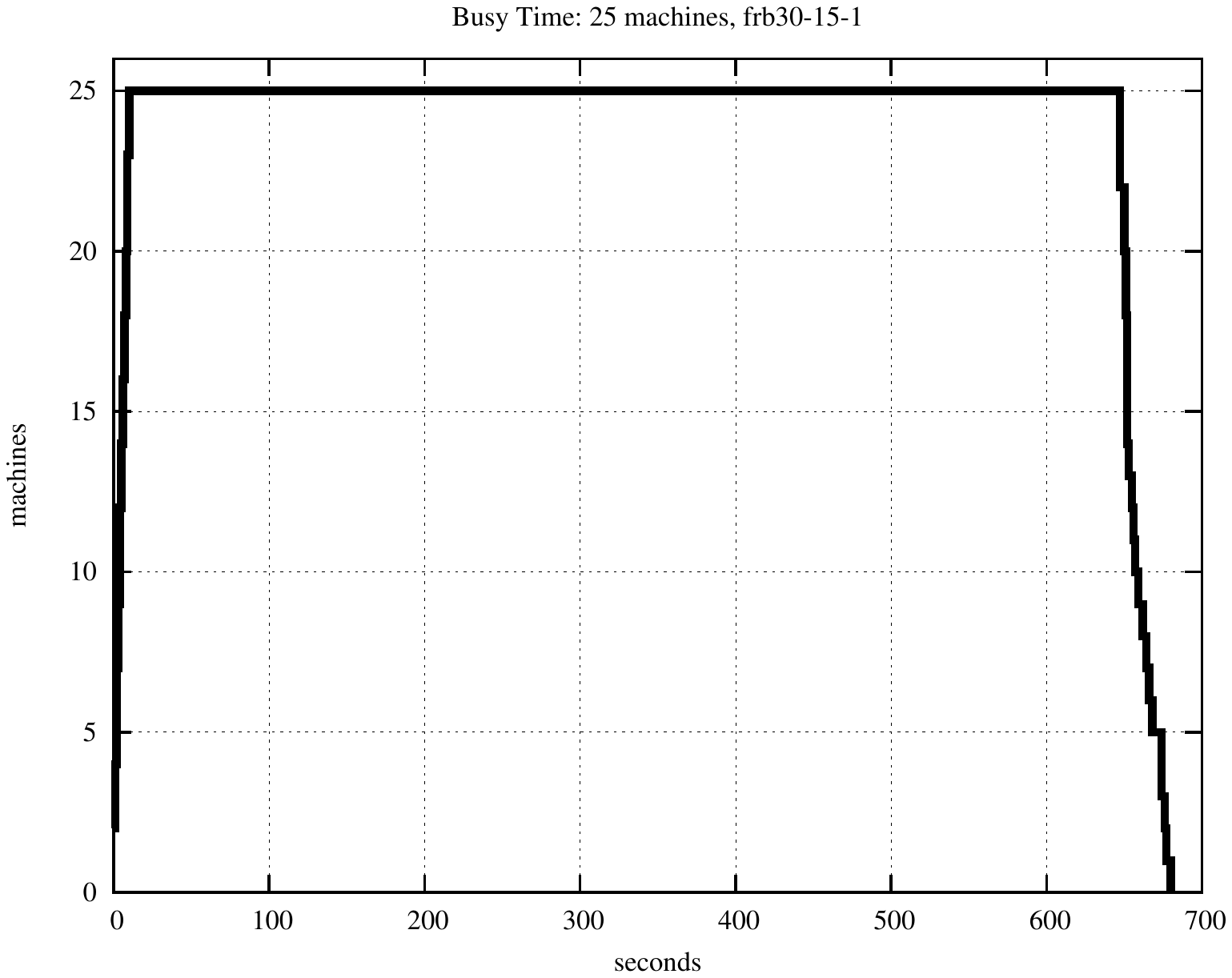}
\end{minipage}
\hfill
\begin{minipage}[t]{0.30\textwidth}
\includegraphics[height=9.0cm]{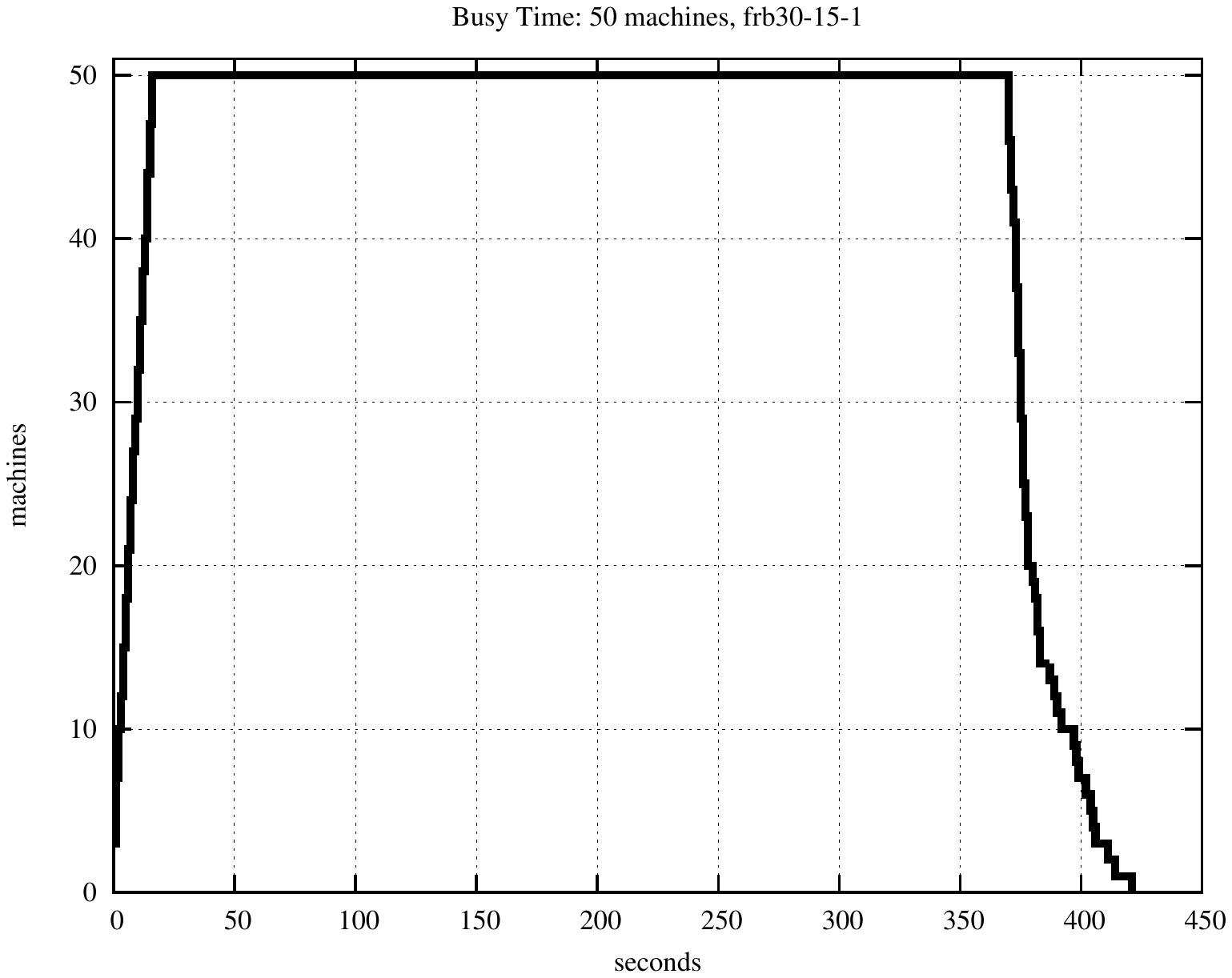}
\end{minipage}
\hfill
\begin{minipage}[t]{0.3\textwidth}
\includegraphics[height=9.0cm]{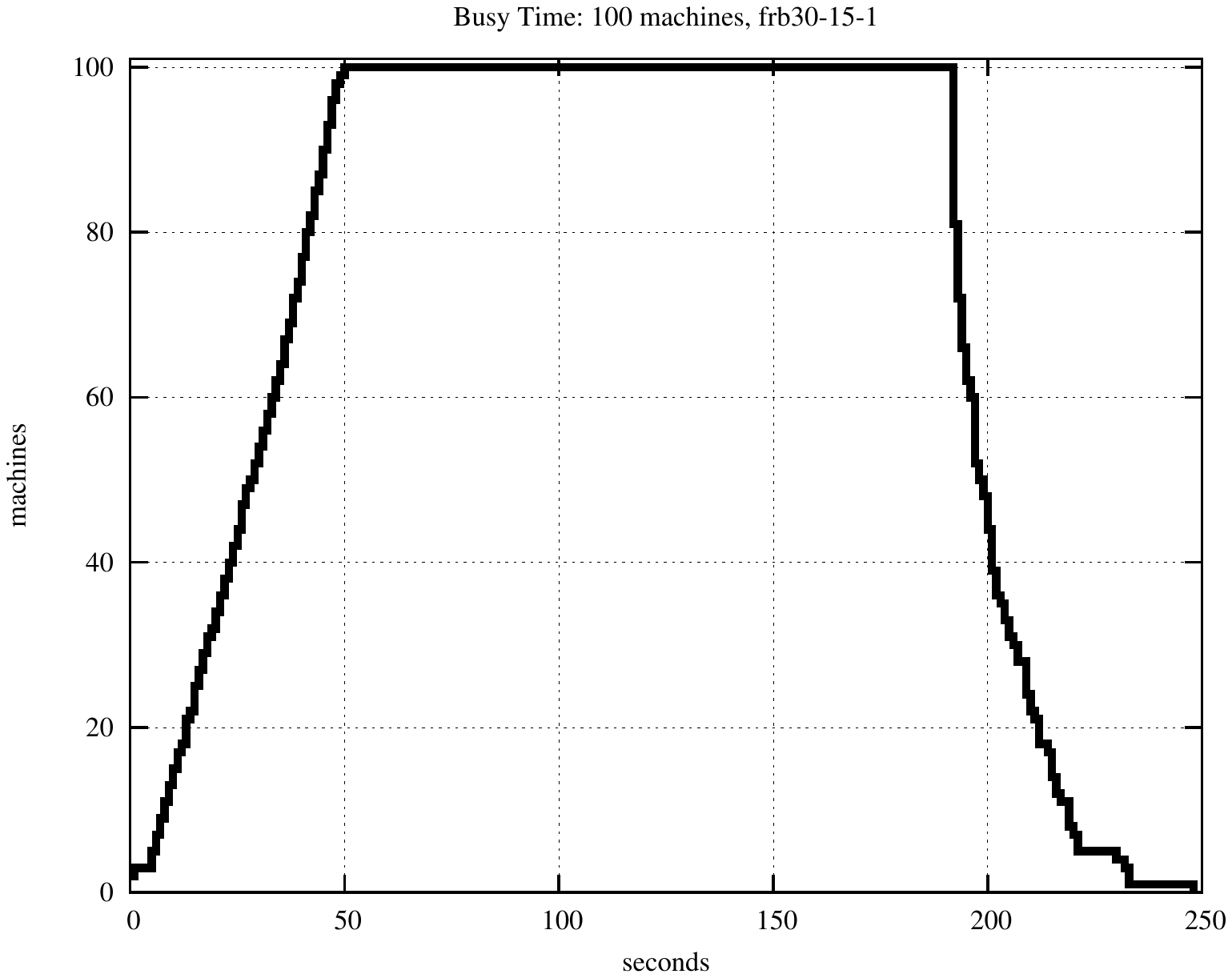}
\end{minipage}
\end{center}
\vspace{-4.0cm}
\begin{center}
\hspace{-1.5cm}
\begin{minipage}[t]{0.3\textwidth}
\includegraphics[height=9.0cm]{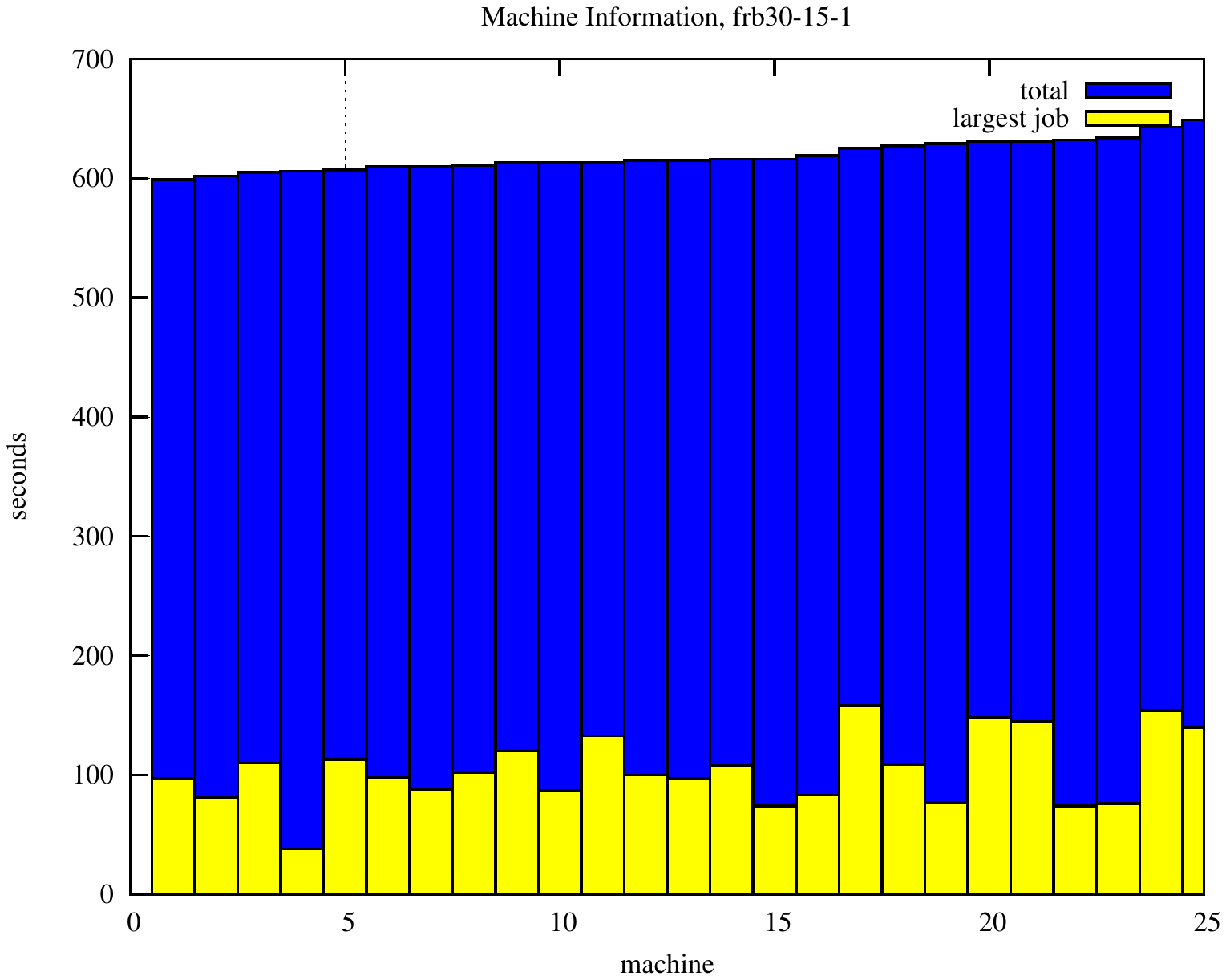}
\end{minipage}
\hfill
\begin{minipage}[t]{0.3\textwidth}
\includegraphics[height=9.0cm]{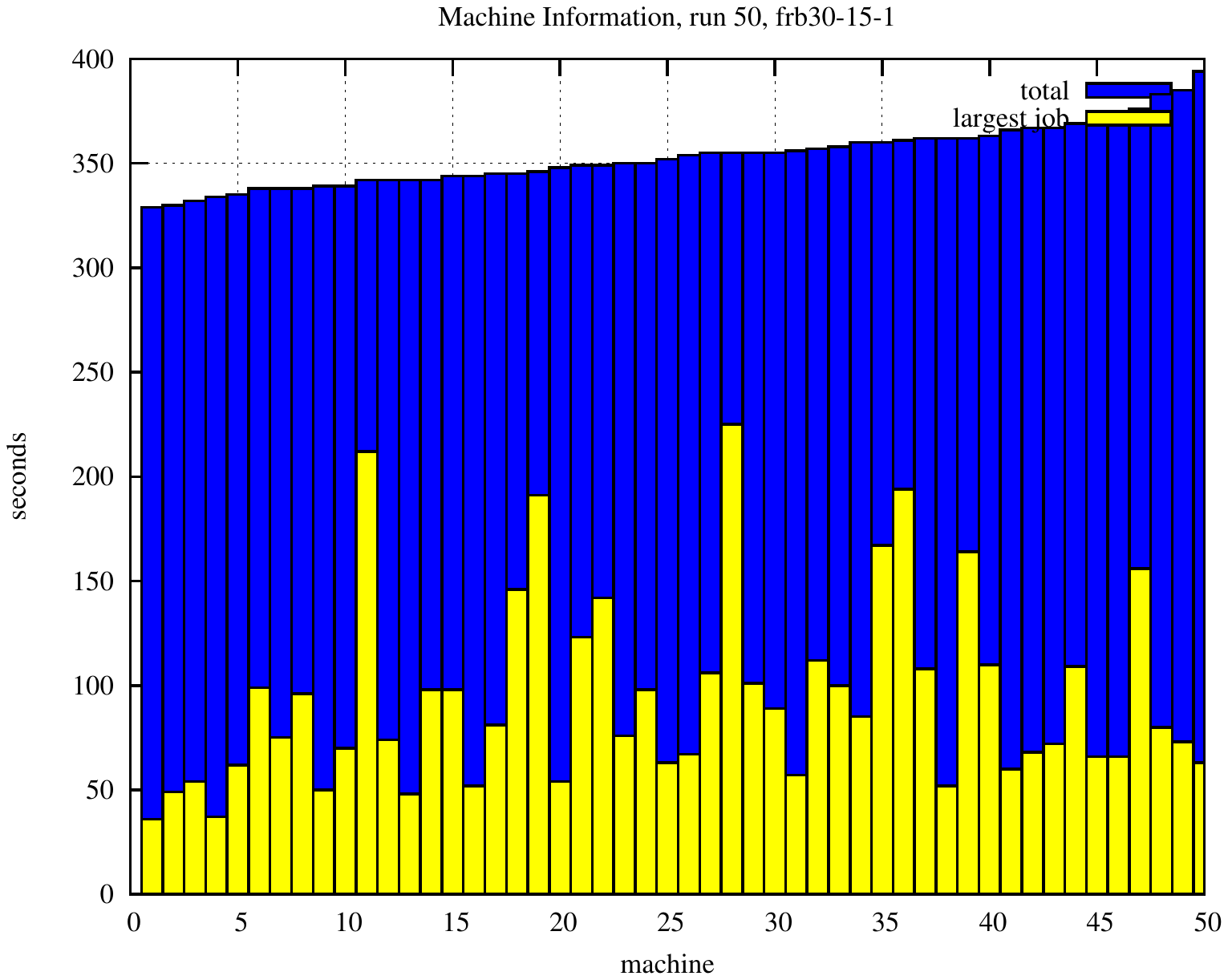}
\end{minipage}
\hfill
\begin{minipage}[t]{0.3\textwidth}
\includegraphics[height=9.0cm]{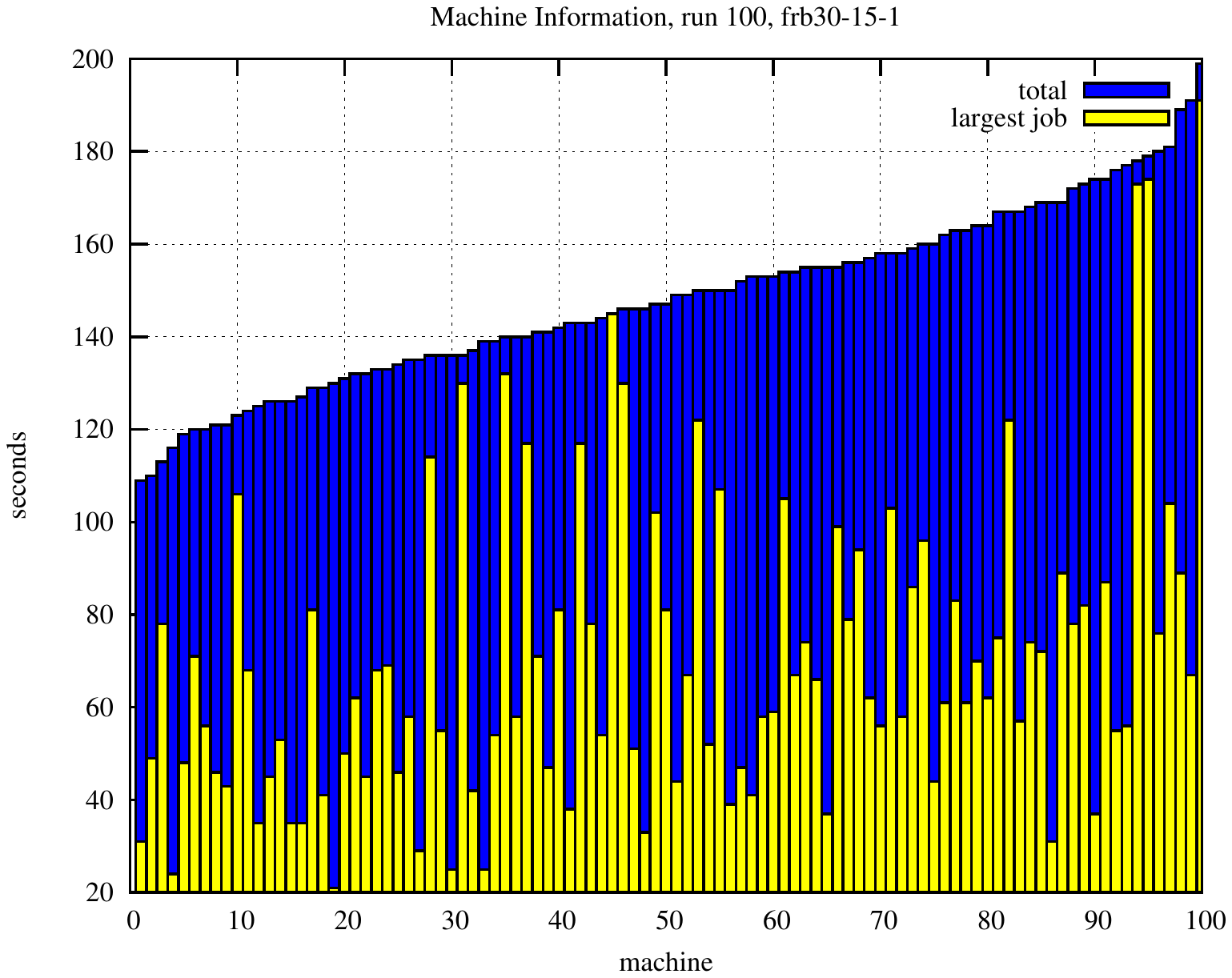}
\end{minipage}
\end{center}
\vspace{-4.0cm}
\begin{center}
\hspace{-1.5cm}
\begin{minipage}[t]{0.3\textwidth}
\includegraphics[height=9.0cm]{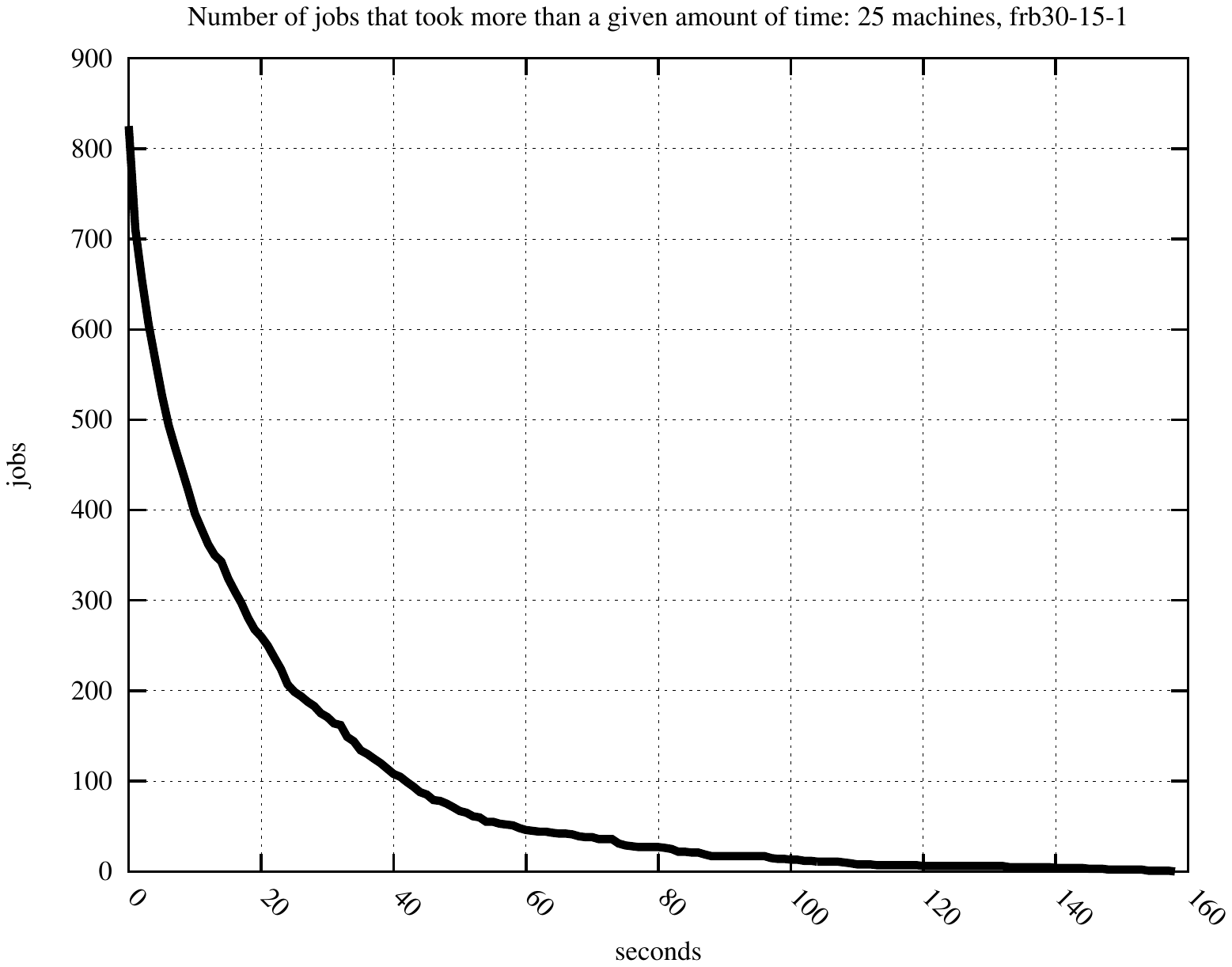}
\end{minipage}
\hfill
\begin{minipage}[t]{0.3\textwidth}
\includegraphics[height=9.0cm]{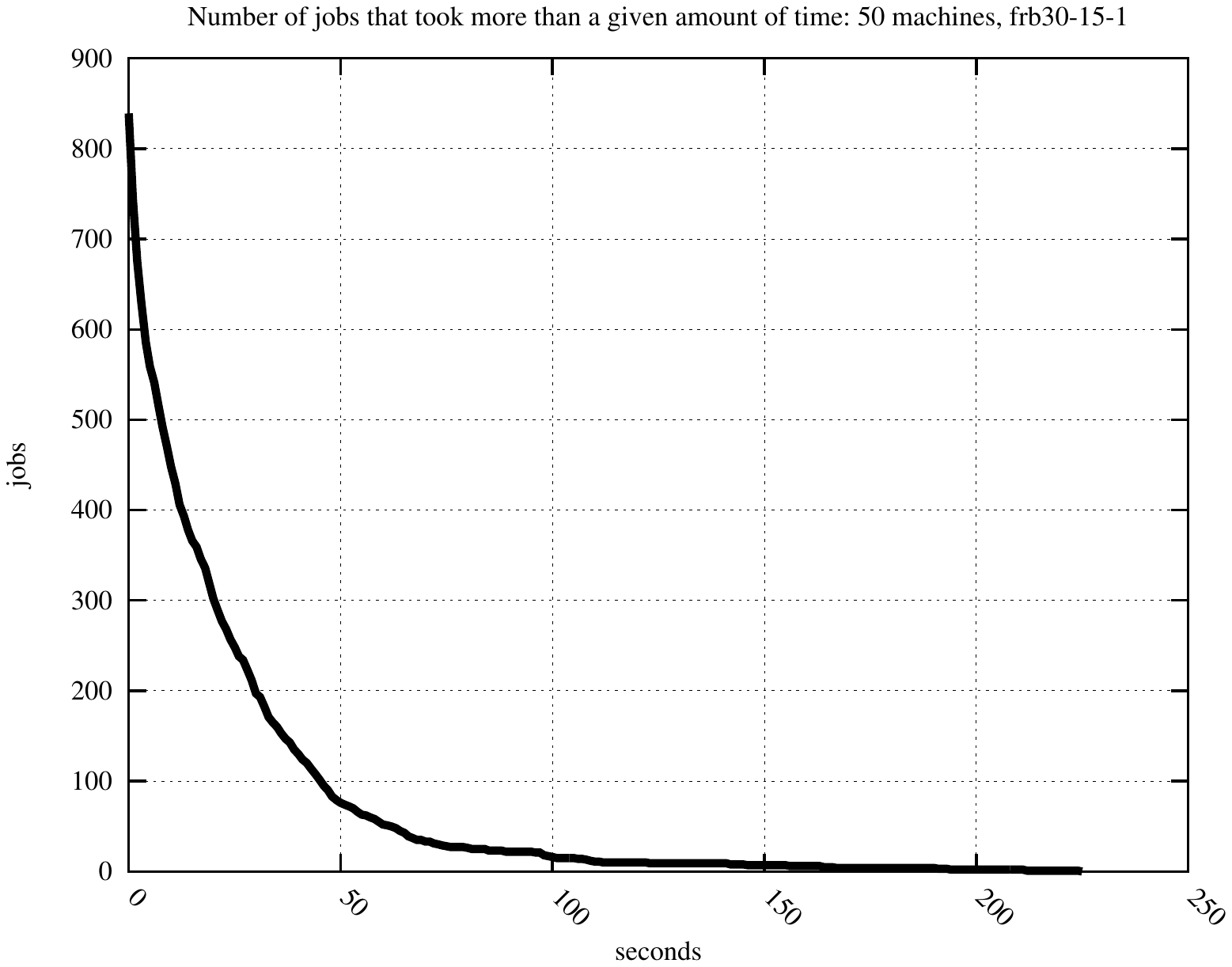}
\end{minipage}
\hfill
\begin{minipage}[t]{0.3\textwidth}
\includegraphics[height=9.0cm]{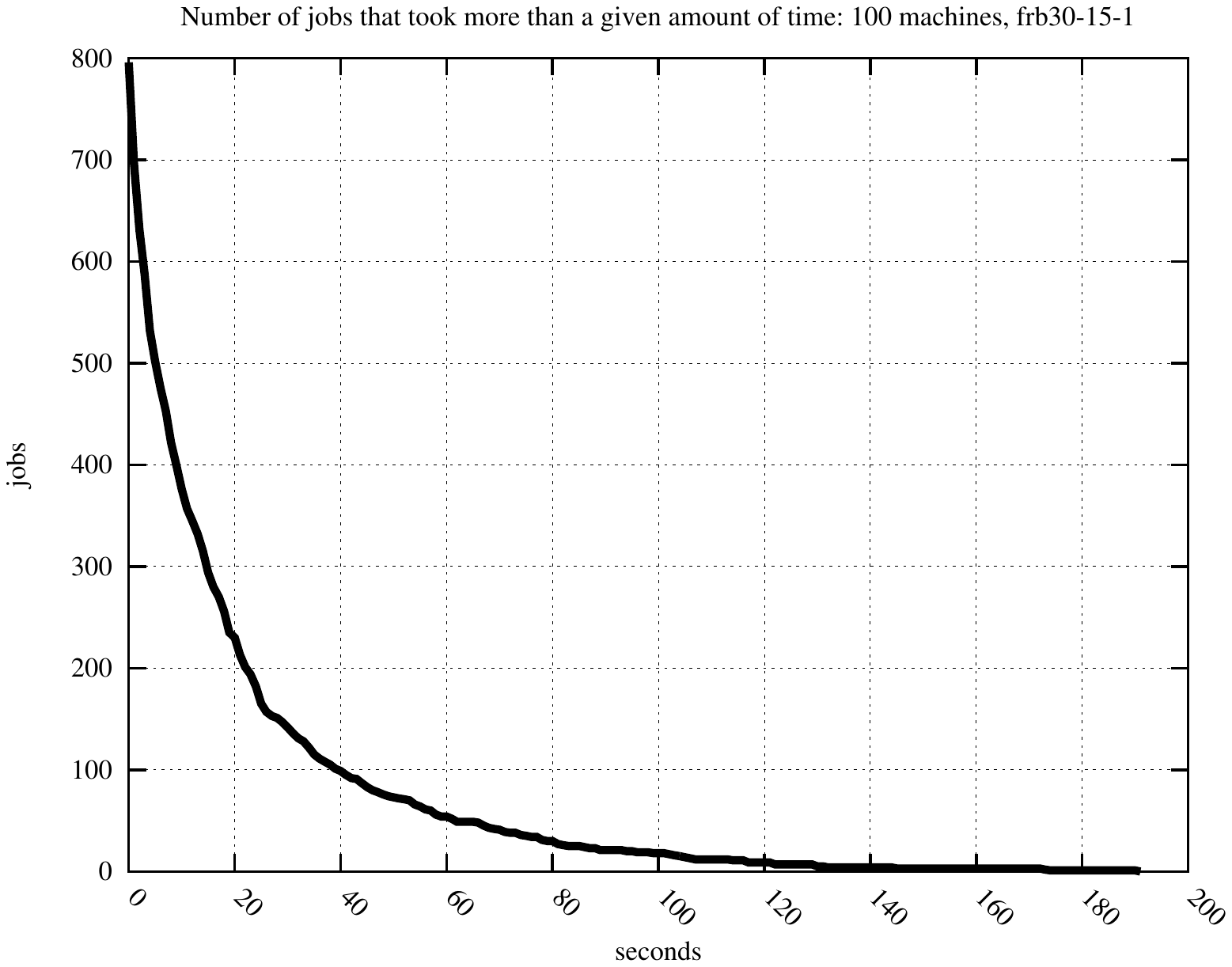}
\end{minipage}
\end{center}
\caption{Instance fr30-15-1 on 25, 50 and 100 machines. Top step graph shows the number of machines busy at any time. Middle histograms shows 
total nodes explored on each machine in blue (sorted in increasing order) and in yellow the number of nodes in longest job. 
Bottom contours show the number of jobs that took more than a given amount of time.}
\label{frb30-15-1}
\end{figure}

Looking at the 25 machine column we see that the worst speed up was 8.0 (frb35-17-5) and the best 39.6 (frb30-15-4) 
and this is a super-linear speed (also seen in p-hat500-3) and should come as no 
surprise \cite{herbSutter,clearwater}\footnote{\cite{clearwater} went so far as to call this a \emph{combinatorial implosion}.}.
This occurs because an early job terminated with a good lower 
bound on the clique size and this
allowed subsequent jobs to terminate quickly. Instance frb30-15-1 is a ``good'' instance, showing an increasing
speedup as we increase the number of machines. This is analysed in Figure \ref{frb30-15-1}.

In Figure \ref{frb30-15-1} step graphs in the top row give the number of machines busy (y-axis) at a given time (x-axis). The first
graph on the left is for the run with 25 machines, the middle for 50 machines and on the right 100 machines. We see a relatively sharp cut-off
with nearly all machines kept busy  right up to termination of the last job. We also see an increasing ramp-up cost for kicking off jobs, 
most notable for 100 machines, taking about 50 seconds to kick off the first 100 jobs, i.e. the start up phase was about 20\% of the total time. 
The three histograms on the middle row give in yellow the length of the longest job on a machine and in
blue the total of the jobs' run times on that machine. The data has been sorted by increasing total run time. For the 25 and 50
machine runs we see that no machine has been dominated by a single long job, whereas in the 100 machine
case we see numerous cases where a single machine was tied up with a single long job. Why is this? When few machines are used
jobs must wait to be kicked off and when they are started they have a new lower bound, whereas in the 100 machine case
it is more likely that a hard job is initiated with a very small initial lower bound and this condemns that job to a long isolated 
execution. The three graphs at the bottom plot the number of jobs (y-axis) that took more than a given amount of time (x-axis).
This shows that the majority of jobs were short (easy) and the minority were long (hard). And this is what we should expect.
Many jobs had small candidate sets and relatively large lower bounds and these terminated quickly. Conversely, there were a few jobs with large
candidate sets and small lower bounds and these are hard. There were also jobs that had a candidate set that contained a largest clique 
or a clique close to that size and it was hard to prove optimality, i.e. where we expect to find the hard problems \cite{ckt,kappa}.

\begin{figure}
\vspace{-3.5cm}
\begin{center}
\hspace{-1.5cm}
\begin{minipage}[t]{0.3\textwidth}
\includegraphics[height=9.0cm]{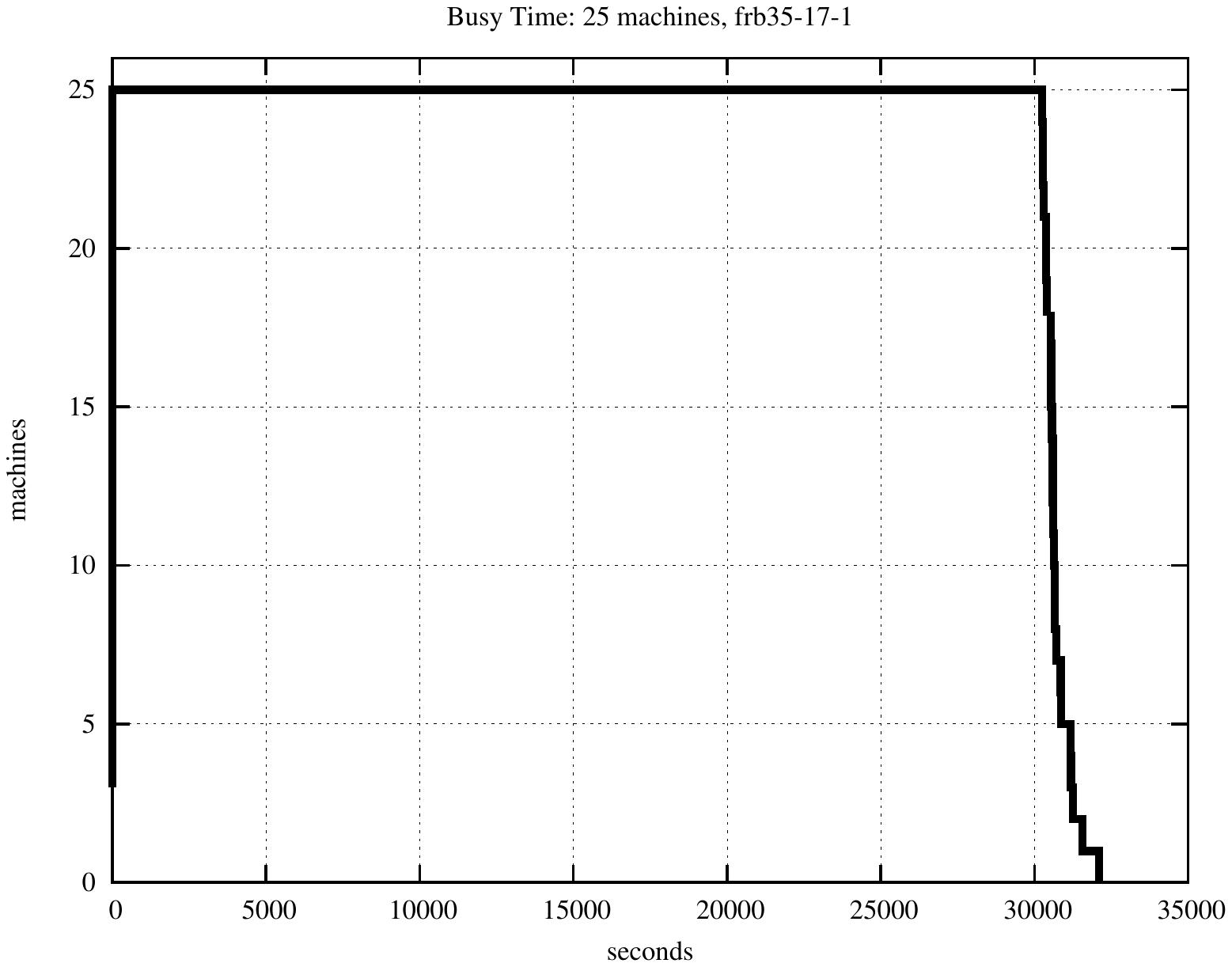}
\end{minipage}
\hfill
\begin{minipage}[t]{0.3\textwidth}
\includegraphics[height=9.0cm]{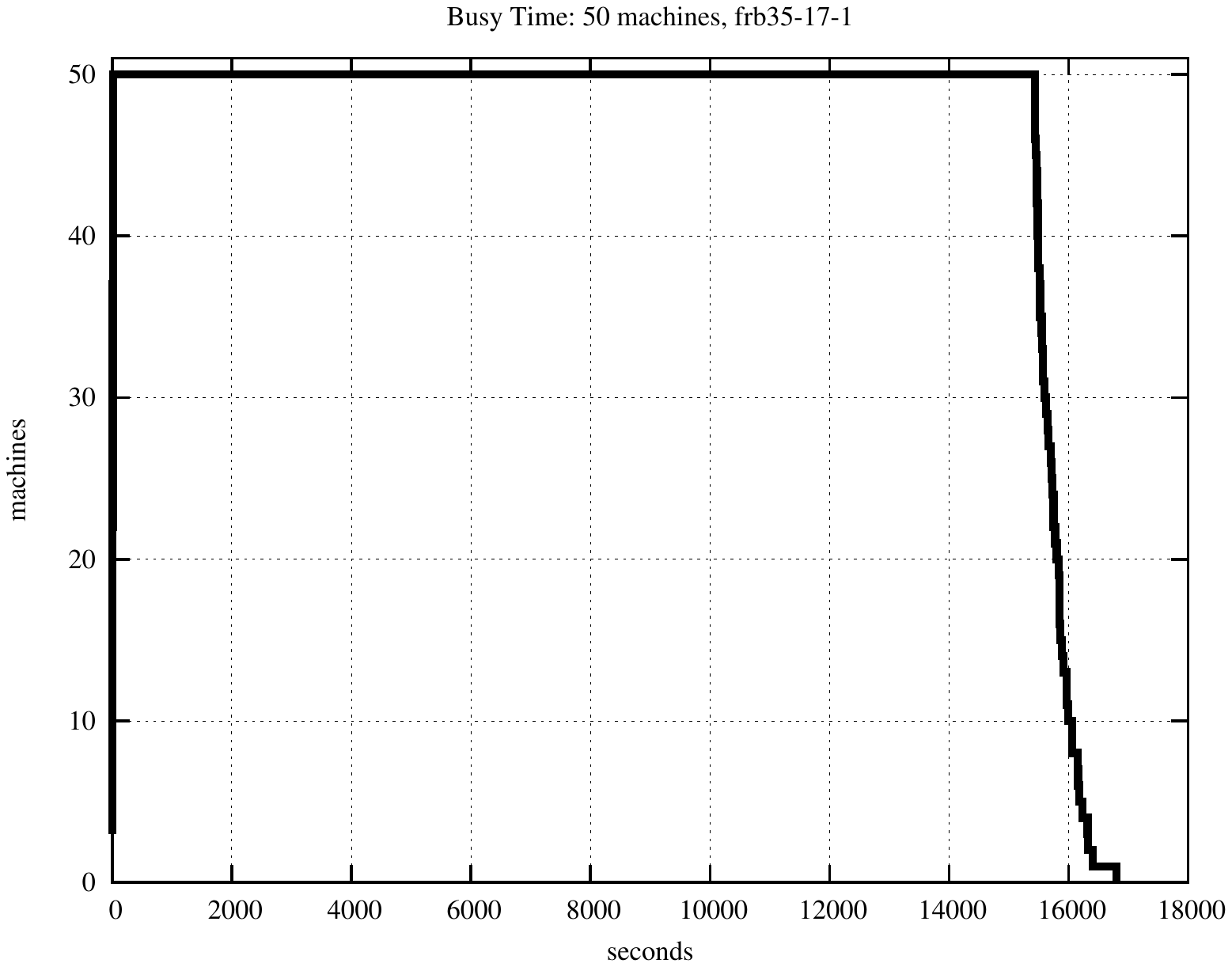}
\end{minipage}
\hfill
\begin{minipage}[t]{0.3\textwidth}
\includegraphics[height=9.0cm]{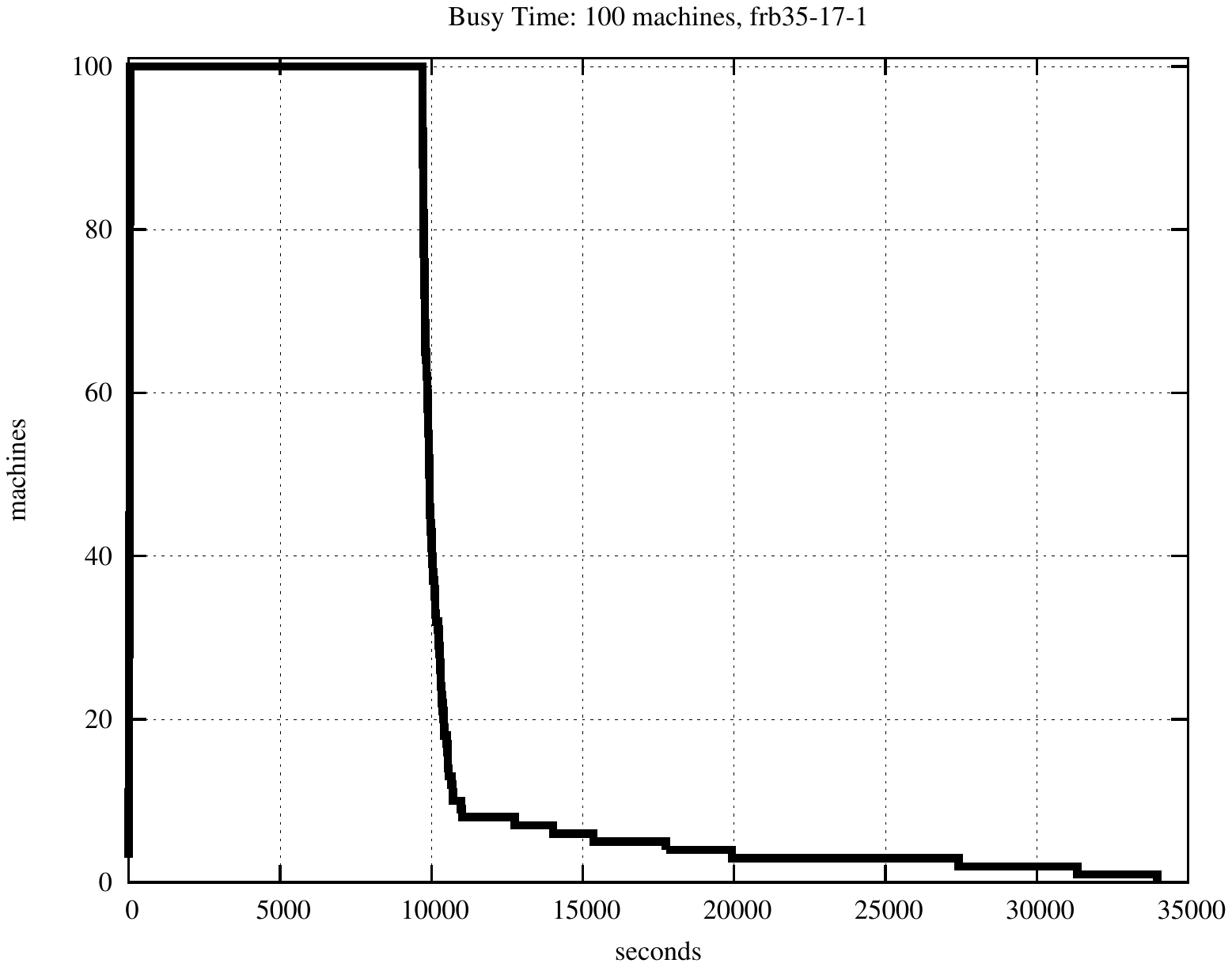}
\end{minipage}
\end{center}
\vspace{-4.0cm}
\begin{center}
\hspace{-1.5cm}
\begin{minipage}[t]{0.3\textwidth}
\includegraphics[height=9.0cm]{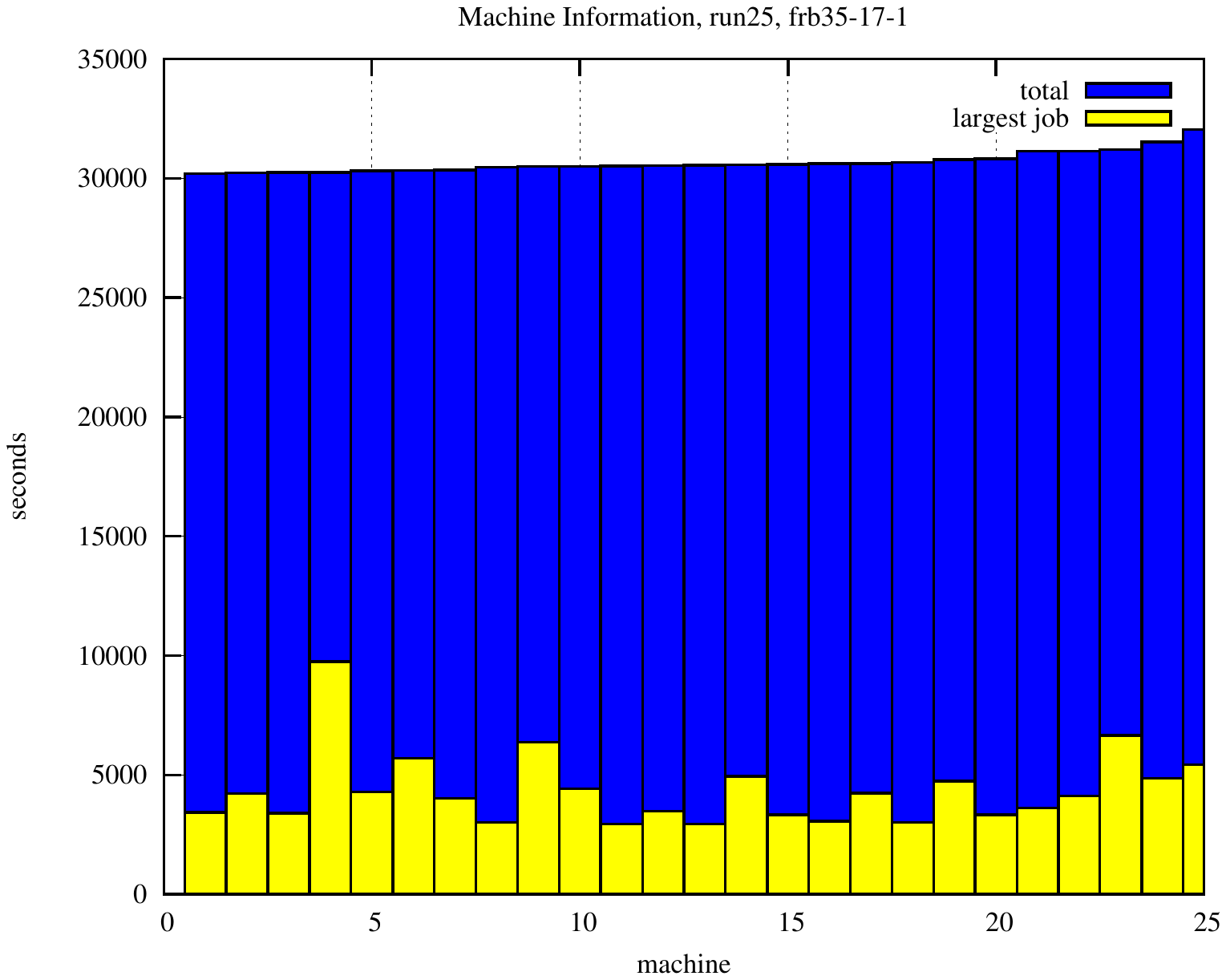}
\end{minipage}
\hfill
\begin{minipage}[t]{0.3\textwidth}
\includegraphics[height=9.0cm]{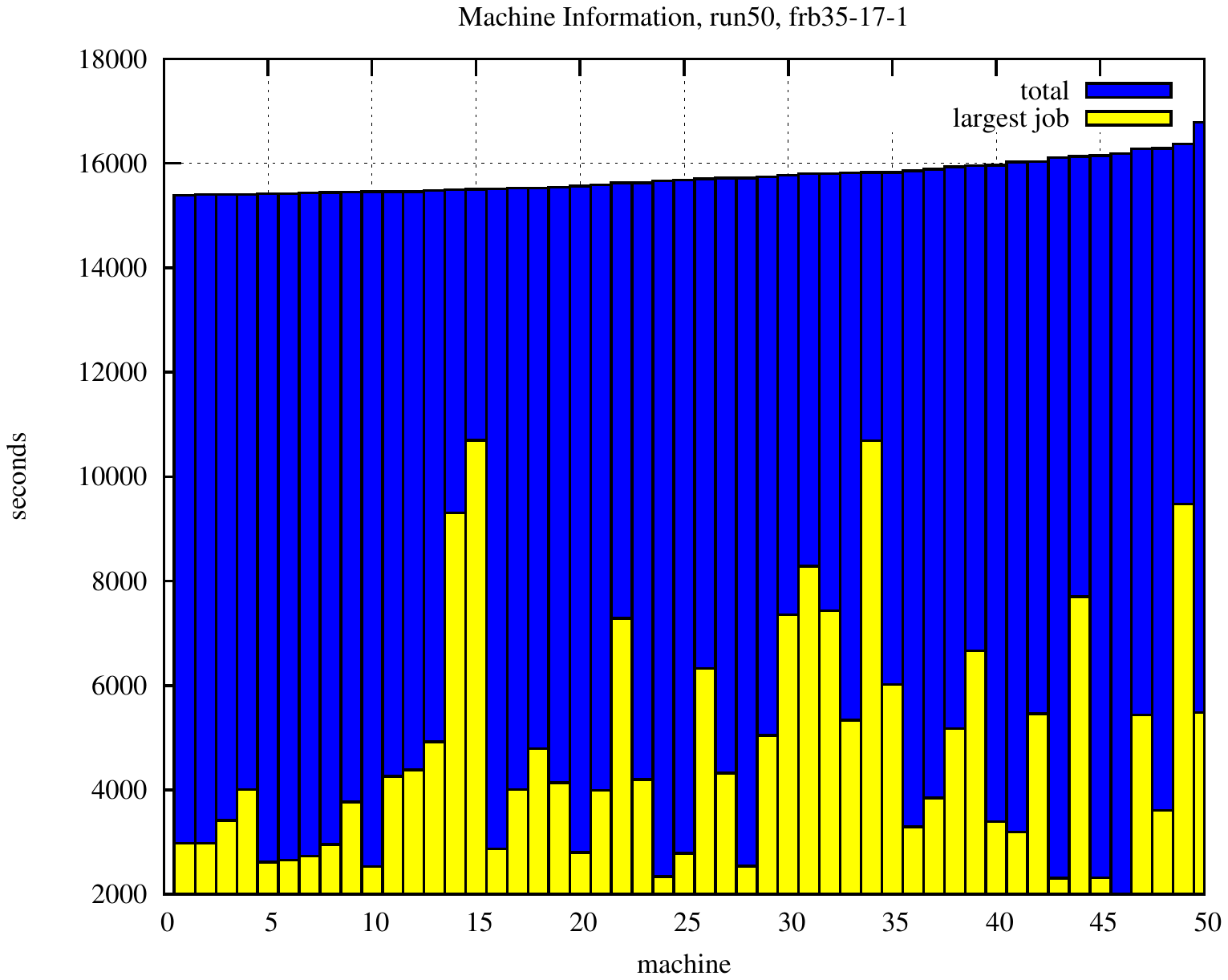}
\end{minipage}
\hfill
\begin{minipage}[t]{0.3\textwidth}
\includegraphics[height=9.0cm]{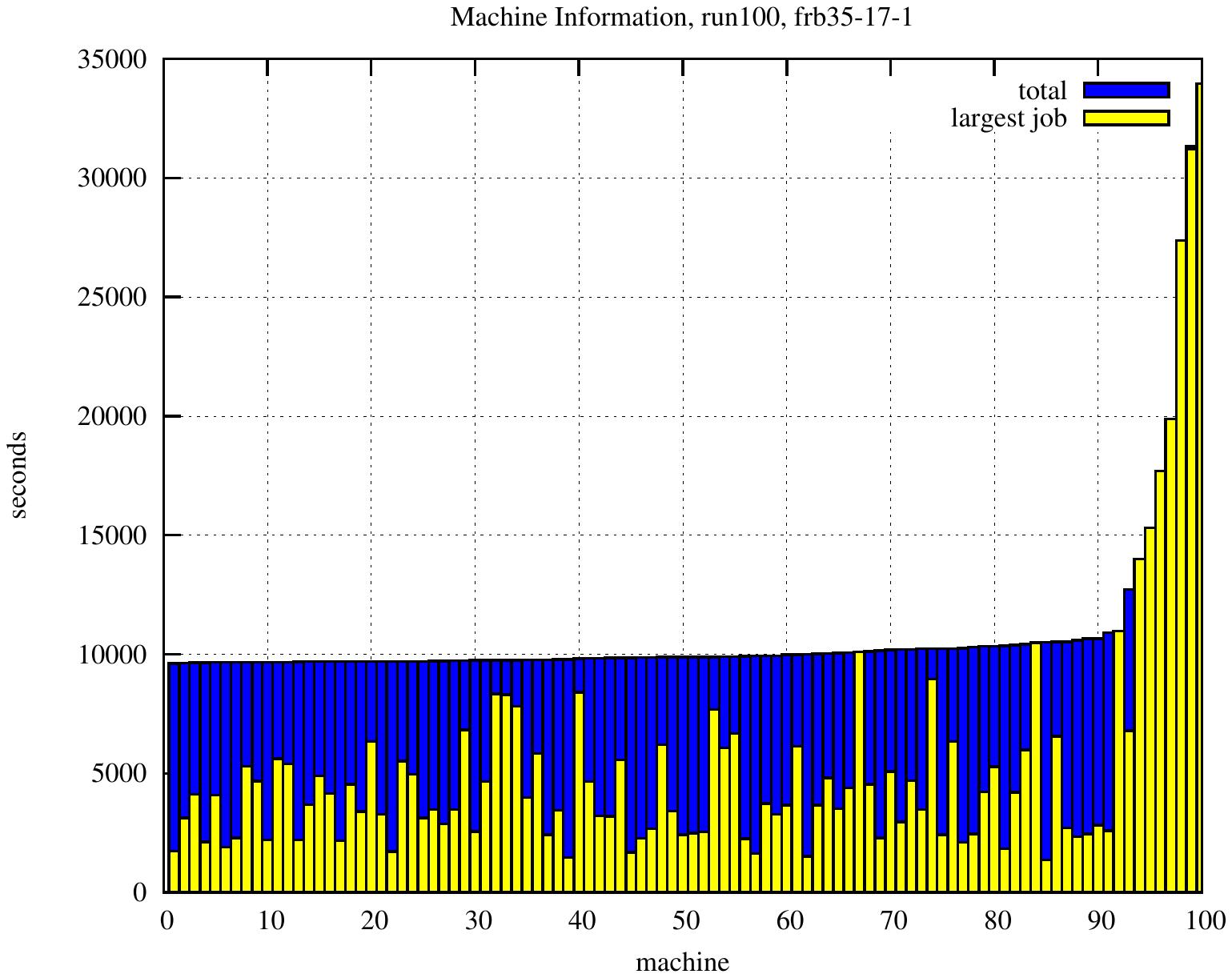}
\end{minipage}
\end{center}
\vspace{-4.0cm}
\begin{center}
\hspace{-1.5cm}
\begin{minipage}[t]{0.3\textwidth}
\includegraphics[height=9.0cm]{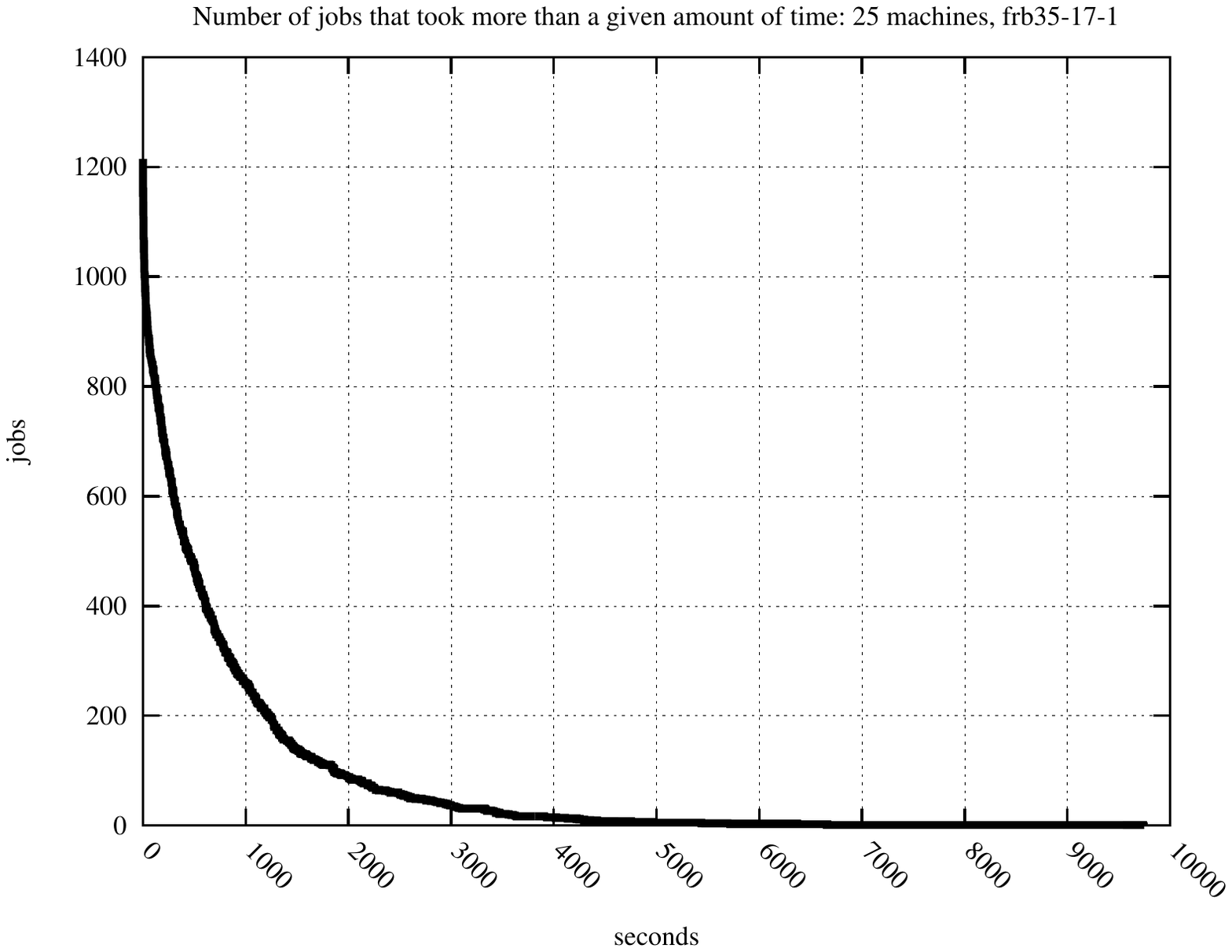}
\end{minipage}
\hfill
\begin{minipage}[t]{0.3\textwidth}
\includegraphics[height=9.0cm]{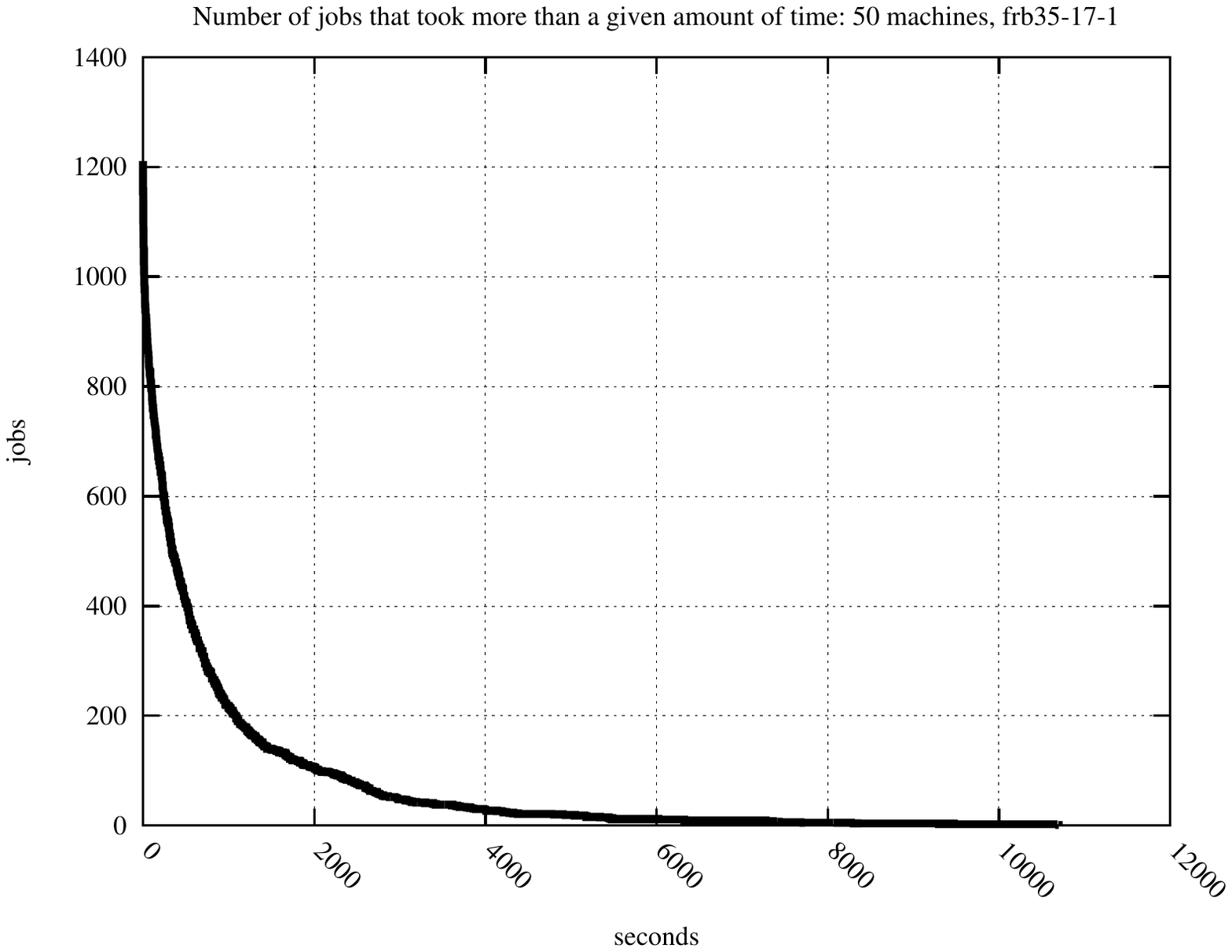}
\end{minipage}
\hfill
\begin{minipage}[t]{0.3\textwidth}
\includegraphics[height=9.0cm]{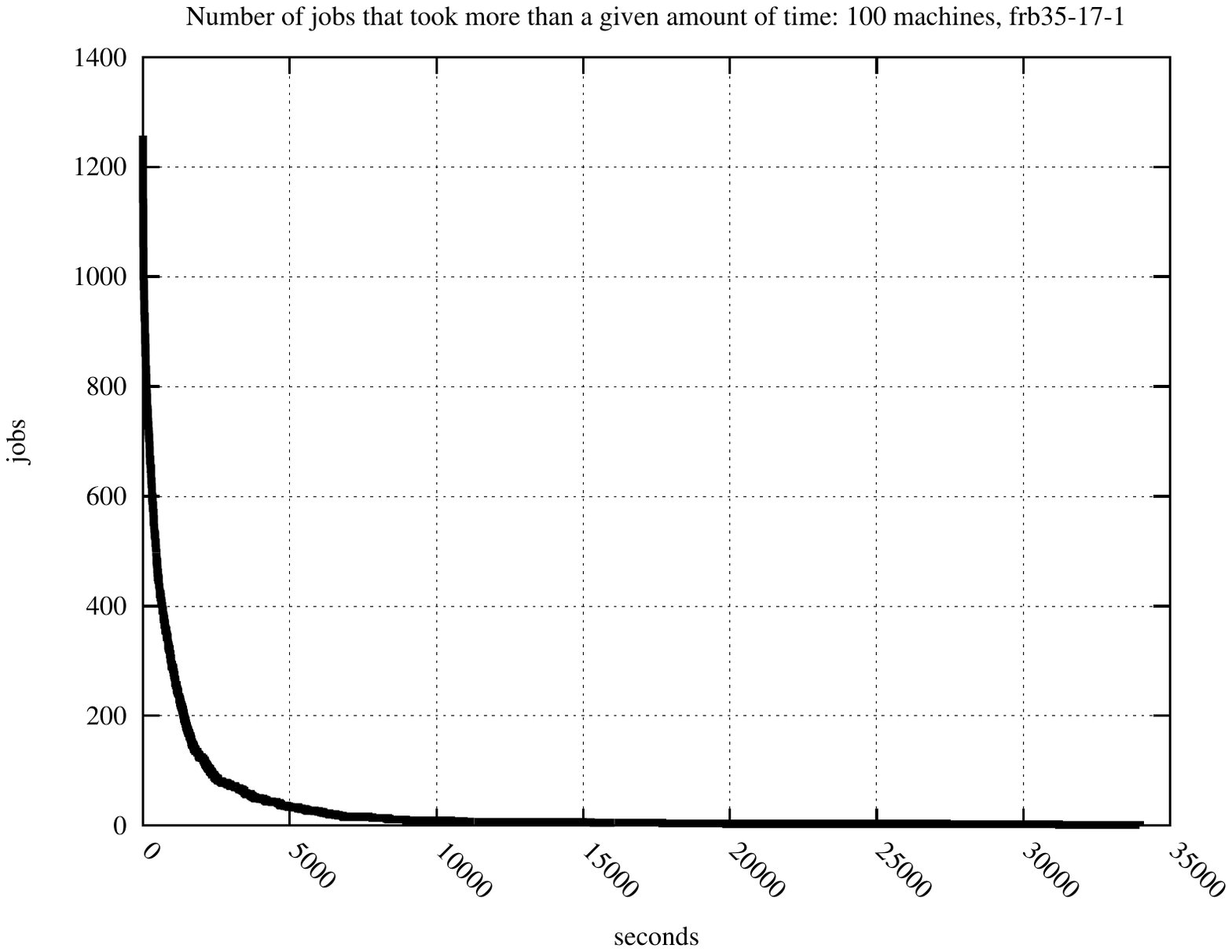}
\end{minipage}
\end{center}
\caption{Instance fr35-17-1 on 25, 50 and 100 machines. Top step graph shows the number of machines busy at any time. Middle histograms shows 
total nodes explored on each machine in blue (sorted in increasing order) and in yellow the number of nodes in longest job. 
Bottom contours show the number of jobs that took more than a given amount of time.}
\label{frb35-17-1}
\end{figure}

Table \ref{bigTable} shows that frb35-17-1 is \emph{not} a ``good'' instance. As we increase the machines from 25 to 50 
speedup improves (from 19.5 to 37.2) but then falls drastically (to 18.4) with 100 machines.
This is analysed in Figure \ref{frb35-17-1}. Looking at the rightmost graphs we see 
at top right a long tail: after 10,000 seconds 90 machines are idle and the remaining 23,000 seconds is taken up by a few
hard jobs. The rightmost (middle) histogram shows why that is so: there are three jobs that exclusively consumed all of the time on
three machines. The jobs forming the long tail were started with ``best
so far'' values of 12, 12 and 14; had these jobs been started with a ``best so
far'' value of 30 or greater (or had they been fed that value later), they would each have terminated in under a second. 

\begin{table}
\begin{center}
\begin{scriptsize}
\begin{tabular}{|c|c|c|c|c|c c|c c|c c|} \hline 
instance & $n$ & $p$ & $\omega$ & $mc_{1}$  & $mc_{25}$ & (gain)  & $mc_{50}$ & (gain)   & $mc_{100}$  & (gain)   \\ \hline
1000-10-00  & 1,000 & 0.1 & 6     & 0          & 266      & (0.0)    & 267      & (0.0)    & 393      & (0.0)    \\
1000-10-01   &&& 6     & 0          & 271      & (0.0)    & 411      & (0.0)    & 331      & (0.0)    \\
1000-10-02   &&& 6     & 0          & 274      & (0.0)    & 261      & (0.0)    & 255      & (0.0)    \\
1000-10-03   &&& 5     & 0          & 248      & (0.0)    & 238      & (0.0)    & 255      & (0.0)    \\
1000-10-04   &&& 5     & 0          & 252      & (0.0)    & 257      & (0.0)    & 353      & (0.0)    \\
1000-10-05   &&& 6     & 0          & 237      & (0.0)    & 261      & (0.0)    & 371      & (0.0)    \\
1000-10-06   &&& 6     & 0          & 233      & (0.0)    & 263      & (0.0)    & 264      & (0.0)    \\
1000-10-07   &&& 6     & 0          & 239      & (0.0)    & 234      & (0.0)    & 266      & (0.0)    \\
1000-10-08   &&& 6     & 0          & 242      & (0.0)    & 259      & (0.0)    & 287      & (0.0)    \\
1000-10-09   &&& 5     & 0          & 250      & (0.0)    & 252      & (0.0)    & 267      & (0.0)    \\
1000-50-00  & 1,000 & 0.5 & 15    & 1,380       & 392      & (3.5)    & 371      & (3.7)    & 246      & (5.6)    \\
1000-50-01   &&& 15    & 1,397       & 411      & (3.4)    & 360      & (3.9)    & 263      & (5.3)    \\
1000-50-02   &&& 15    & 1,489       & 421      & (3.5)    & 285      & (5.2)    & 276      & (5.4)    \\
1000-50-03   &&& 15    & 1,406       & 372      & (3.8)    & 324      & (4.3)    & 384      & (3.7)    \\
1000-50-04   &&& 15    & 1,491       & 417      & (3.6)    & 304      & (4.9)    & 370      & (4.0)    \\
1000-50-05   &&& 15    & 1,476       & 316      & (4.7)    & 358      & (4.1)    & 397      & (3.7)    \\
1000-50-06   &&& 15    & 1,415       & 283      & (5.0)    & 401      & (3.5)    & 325      & (4.4)    \\
1000-50-07   &&& 15    & 1,430       & 292      & (4.9)    & 373      & (3.8)    & 352      & (4.1)    \\
1000-50-08   &&& 15    & 1,458       & 348      & (4.2)    & 322      & (4.5)    & 579      & (2.5)    \\
1000-50-09   &&& 15    & 1,438       & 334      & (4.3)    & 812      & (1.8)    & 333      & (4.3)    \\
1000-60-00   & 1,000 & 0.6 & 19    & 58,287      & 2,425     & (24.0)   & 1,959     & (29.8)   & --- & (--)   \\
1000-60-01   &&& 19    & 64,421      & 2,710     & (23.8)   & 1,544     & (41.7)   & --- & (--)   \\
1000-60-02   &&& 20    & 49,135      & 2,486     & (19.8)   & 1,144     & (43.0)   & --- & (--)   \\
1000-60-03   &&& 19    & 68,230      & 2,842     & (24.0)   & 1,646     & (41.5)   & --- & (--)   \\
1000-60-04   &&& 19    & 59,667      & 2,524     & (23.6)   & 1,596     & (37.4)   & --- & (--)   \\
1000-60-05   &&& 19    & 65,670      & 2,847     & (23.1)   & 1,554     & (42.3)   & --- & (--)   \\
1000-60-06   &&& 19    & 63,603      & 2,807     & (22.7)   & 1,879     & (33.8)   & --- & (--)   \\
1000-60-07   &&& 20    & 45,740      & 2,213     & (20.7)   & 1,557     & (29.4)   & --- & (--)   \\
1000-60-08   &&& 19    & 61,185      & 2,919     & (21.0)   & 1,469     & (41.7)   & --- & (--)   \\
1000-60-09   &&& 19    & 63,723      & 2,700     & (23.6)   & 1,749     & (36.4)   & --- & (--)   \\ \hline

\end{tabular}
\end{scriptsize}
\end{center}
\caption{Random instance: run time in seconds, using 1 to 100 machines.}
\label{randomTable}
\end{table}

\subsection{Random Instances}
Table \ref{randomTable} gives results on 30 random graphs $G(n,p)$ all with $n = 1000$ vertices. The first 10 graphs have
edge probability of $p = 0.1$, $p = 0.5$ for the next 10 graphs and $p = 0.6$ for the last10. 
The $G(1000,0.1)$ instances are easy on a single processor,
taking less than a second. Therefore the run time on many machines is largely overhead of dispatching $n \times 8 = 8,000$ jobs and collecting 
their results. This demonstrates that our approach is only applicable to large hard instances where that overhead can be amortised.
The $G(1000,0.5)$ instances are relatively hard (about 20 minutes on a single machine) and the overhead pays off modestly
with a speedup between 1.8 (1000-50-09 with 50 machines) and 5.6 (1000-50-00 with 100 machines). Why so modest? Analysing instance
1000-50-00 shows that none of the jobs took more than two seconds runtime yet incurred the same overhead as the easier $G(1000,0.1)$
instances. So again, the overhead remains significant for these instances. For $p = 0.6$ only 50 machines were available to us,
nevertheless we see speedups from a minimum of 19.8 (25 machines) up to 42.3 (50 machines).

\subsection{Fault Tolerance}
On several occasions during execution of some of the larger problems, a machine
was either powered off or rebooted. Here the NFS implementation was an
advantage: the lost task could easily be identified and restarted. The
``results'' directory also provides an easy check that every task was in fact
executed. The system is in no way tolerant of failures of the NFS server, but experience
suggests that the NFS server is far more reliable than the lab machines.

\section{Potential Improvements}
\label{sec:improvements}
\vspace{-1.5mm}
On reflection, there are many things we could do to improve performance if we were given more time.

\paragraph{Order of Task Execution:}
We use a random order of task execution as a simple way of reducing contention.
This is unlikely to be optimal, as the initial ordering of vertices has a large
effect upon performance. This also significantly affects reproducibility of
results.  Although reproducibility could be improved by using fixed per-machine
orderings, a more sophisticated implementation that did not have to rely upon
NFS would be able to dispatch jobs in the order in which they would be executed
in the non-parallel version of the algorithm.

\paragraph{Slow Startup:}
With the current implementation it can take up to a minute for all 100 machines
to start running. This is because SSH login attempts are deliberately rate
limited. We could avoid this cost with a more sophisticated startup mechanism.

\paragraph{Re-reading the Best So Far:}
We only read and write the file containing the best clique found so far at the
start and end of a task respectively. It would be better to do this more
frequently. However, due to locking, this has considerable overhead on NFS.
This also affects reproducibility of results: in some cases, a small delay
before starting a job would lead to it having a better ``best so far'' value,
which in turn would vastly reduce runtimes. 
However, some of the long tails in other problems would not be removed by this
method. In some cases, a better initial ``best so far'' is of no help for the
most time-consuming subproblems.

\paragraph{Finer Granularity vs Work-Stealing:}
To reduce the long tail in cases where re-reading ``best so far'' does not
help, we could split the second level fully into (less than) $m^2$ jobs, or
split on the first three levels. This was not possible with the NFS server
available to us, but would be an option for better implementations.
An increase in splitting is still not enough to remove every long tail,
however. In some cases there are a small number of areas deep down in the tree
that contribute most to the runtime. Dealing with these would need some kind of
work stealing mechanism, which in turn requires much more sophisticated
communication than was available.

\paragraph{Threading the Workers:}
Each of the lab machines we had available was dual core. Memory limitations
prevented us from running two instances of the worker program per machine;
threading the client (and adding a second set of in-process locking, since file
locks are per-program rather than per-thread) could possibly give us the
equivalent of doubling the number of machines, at the expense of a more
complicated implementation.

\paragraph{Not Using Java:}
Java was used due to an existing implementation being available, and because of
the ease of running Java programs on non-identical systems. A reimplementation
in a faster language and the use of bit encoded sets (such as in \cite{segundo2011}) would produce a substantial
speed-up, at the cost of increased development time and complexity of
implementation. 

\section{Conclusion}
So, how did we do? Did we get a \emph{costup}? Did we get an increase in performance greater than the increase in cost?
We think so. Just looking at the frb35 instances, each instance typically takes weeks to solve on a single machine. We solve frb35-17-5 in
11 hours, frb35-17-2 in under 7 hours (and over 13 days on a single machine), frb35-17-1 in about 9 hours 
(and more than a week on a single machine),
frb35-17-4 in under 5 hours and frb35-17-3 in under 4 hours. We have spent less
than a week to do this, less than a week to get more than a week's speedup.

Many of our difficulties were down to performance limitations with NFS, 
and these would be vastly reduced if we were using a shared memory multi-core or even just a NUMA multi-processor system. 
In other words, we believe that things are going to get better for this kind of costup in the future, not worse.

\end{document}